\documentclass[12pt,preprint]{aastex}
\begin{document}

\title{On the Relaxation Behaviors of Slow and Classical Glitches: Observational Biases and Their Opposite Recovery Trends }
\author{Yi Xie\altaffilmark{1,2}, Shuang-Nan Zhang\altaffilmark{1,3}}
\altaffiltext{1}{National Astronomical Observatories, Chinese
Academy of Sciences, Beijing, 100012, China}

\altaffiltext{2}{University of Chinese Academy of Sciences, Beijing,
100049, China}

\altaffiltext{3}{Key Laboratory of Particle Astrophysics, Institute
of High Energy Physics, Chinese Academy of Sciences, Beijing 100049,
China; zhangsn@ihep.ac.cn}

\begin{abstract}

We study the pulsar timing properties and the data analysis methods
during glitch recoveries. In some cases one first fits the
time-of-arrivals (TOAs) to obtain the ``time-averaged" frequency
$\nu$ and its first derivative $\dot\nu$, and then fits models to
them. However, our simulations show that $\nu$ and $\dot\nu$
obtained this way are systematically biased, unless the time
intervals between the nearby data points of TOAs are smaller than
about $10^4$~s, which is much shorter than typical observation
intervals. Alternatively, glitch parameters can be obtained by
fitting the phases directly with relatively smaller biases; but the
initial recovery timescale is usually chosen by eyes, which may
introduce a strong bias. We also construct a phenomenological model
by assuming a pulsar's spin-down law of $\dot\nu\nu^{-3} =-H_0 G(t)$
with $G(t)=1+\kappa e^{-t/\tau}$ for a glitch recovery, where $H_0$
is a constant and $\kappa$ and $\tau$ are the glitch parameters to
be found. This model can reproduce the observed data of slow
glitches from B1822--09 and a giant classical glitch of B2334+61,
with $\kappa<0$ or $\kappa>0$, respectively. We then use this model
to simulate TOA data and test several fitting procedures for a
glitch recovery. The best procedure is: 1) use a very high order
polynomial (e.g. to 50th order) to precisely describe the phase; 2)
then obtain $\nu(t)$ and $\dot\nu(t)$ from the polynomial; and 3)
the glitch parameters are obtained from $\nu(t)$ or $\dot\nu(t)$.
Finally, the uncertainty in the starting time $t_0$ of a classical
glitch causes uncertainties to some glitch parameters, but less so
to a slow glitch and $t_0$ of which can be determined from data.

\end{abstract}


%%%%%%%%%%%%%%%%%%%%%%%%%%%%%%%%%%%%%%%%%%%%%%%%%%%%%%%%%%%%%%%%
\section{Introduction}           %% first-level sections will be auto-capitalized
\label{sect:intro}

Pulsars are very stable rotators. However, many pulsars exhibit
significant timing irregularities, i.e., unpredicted arrival times
of pulses. There are two main types of timing irregularities, namely
`timing noise' which is consisted of low-frequency quasi-periodic
structures, and `glitches' which are abrupt increases in their spin
rates followed by relaxations.

Glitch activities are more frequent in relatively young pulsars with
a characteristic age of $10^4-10^5$~yr (Shemar \& Lyne 1996; Wang et
al. 2000). For the hundreds of glitches
observed\footnote{http://www.atnf.csiro.au/people/pulsar/psrcat/glitchTbl.html.},
their typical fractional jumps in spin frequency $\nu$ are in the
range of $\Delta\nu/\nu\approx10^{-11}-10^{-5}$, and their relative
increment in frequency derivative is
$\Delta\dot{\nu}/\dot{\nu}\sim10^{-3}$. Despite the abundance of
observational data accumulated for over $40$ years, we are still far
from satisfactory understanding of glitch events. Traditional models
mainly involve the expected superfluid nature of part of the neutron
star interior (Anderson \& Itoh 1975; Ruderman 1976), and the
angular momentum is carried in the form of microscopic, quantized
vortices, whose density determines the rotation rate of a pulsar.
Mostly, these vortices are pinned to the crust and the charged
matter in the core of the star, thus their outward drifting motions
are prevented (Anderson \& Itoh 1975; Alpar 1977; Pines et al. 1980;
Alpar et al. 1981; Anderson et al. 1982). However, as the crust
spins down due to the electromagnetic braking, a rotational lag and
stress (Magnus force) gradually builds up. A glitch occurs when the
stress reaches some critical value and the pinning breaks, vortices
suddenly move outward and impart their angular momentum to the
crust. Immediately after the glitch, the vortices are pinned to
other parts again and the superfluid is effectively decoupled from
the crust.

Following the seminal work of Baym, Pethick \& Pines (1969), there
are two classes of models that have been developed to explore the
dynamical evolution of pinned superfluid during the post-glitch
recovery. One kind of models involve a weak coupling between the
superfluid and the crust due to the interaction between free
vortices and the coulomb lattice of nuclei (Jones 1990, 1992, 1998).
Another kind of models assume that the vortices creep rate is highly
temperature-dependent. As the vortices creep through the crust,
angular momentum is gradually transferred (Alpar 1984a, 1984b; Link,
Epstein \& Baym 1993; Larson \& Link 2002). Superfluid vortex
dynamics can model the relaxation well; however, there are still
many significant problems unsolved. For instance, the mechanism that
triggers the glitch in the first place and the detailed processes of
angular momentum transfer during the recovery are still
controversial. It has been suggested that such an event may be
triggered by large temperature perturbations (Link \& Epstein 1996),
or caused by starquakes (Baym \& Pines 1971; Cheng et al. 1992), or
the interactions of the proton vortices and the crustal magnetic
field (Sedrakian \& Cordes 1999), or the superfluid r-mode
instability (Andersson, Comer, \& Prix 2003; Glampedakis \&
Andersson 2009).

Very recently, Pizzochero (2011) proposed an analytic model for
angular momentum transfer associated with Vela-like glitches for the
storage and release of superfluid vorticity, and Seveso et al.
(2012) and Haskell et al. (2013) extended the model to realistic
equations of state and relativistic backgrounds. Haskell et al.
(2012) further modeled all stages of Vela glitches with a two-fluid
hydrodynamical approach. Furthermore, Haskell \& Antonopoulou (2013)
showed that if glitches are indeed due to large scale unpinning of
superfluid vortices, the different regions in which the unpinning
occurs and the respective timescales on which they recouple can lead
to various observed jump and relaxation signatures. However, by
combining the latest observational data for prolific glitching
pulsars with theoretical results for the crust entrainment,
Andersson et al. (2012) found that the required superfluid reservoir
exceeds that available in the crust. Coincidentally, Chamel (2013)
found that the glitches observed in the Vela pulsar require an
additional reservoir of angular momentum, since the maximum amount
of angular momentum that can possibly be transferred during glitches
is severely limited by the non-dissipative entrainment effects. This
challenges superfluid vortex model of the glitch phenomenon.
Besides, some of the glitch events, such as those with persistent
offset in the spin-down rate of the Crab pulsar following the 1975
glitch is difficult to explain with the dynamic coupling between the
crust and the superfluid interior. An alternative explanation of the
observed frequency deficit is an increase in the external torque
caused by a rearrangement of the stellar magnetic field (Link 1992,
1998). Observationally, many pulsar phenomena, including the mode
changing, pulsar-shape variability and spin-down rates switching,
are caused by changes in pulsar's magnetosphere (Lyne et al. 2010).
Thus these relaxation processes may also be produced by the
magnetosphere activities, which are induced by initial starquakes.

It has also been observed in recent years that some pulsars (e.g.
PSR J1825-0935 and PSR J1835-1106) show another type of irregularity
characterized by a gradual increase in $\nu$, accompanied by a rapid
decrease in $|\dot\nu|$ and subsequent exponential increase back to
its initial value (Zou et al. 2004; Shabanova 2005). That is the
so-called `slow glitch'. Currently, there is still no convincing
theoretical understanding for slow glitches. Peng \& Xu (2008)
proposed that, after a collapse or a small star-quake, the solid
superficial layer of a rigid quark star may be heated and becomes a
viscous fluid, which will eventually produce a gradual increase in
$\nu$. However, Hobbs et al. (2010) and Lyne et al. (2010) argued
that the slow glitches have the same origin as the timing noise of
many pulsars.

For the recovery processes of both glitch and slow glitch events,
the variations of spin frequency $\nu$ and its first derivative
$\dot{\nu}$ of pulsars are obtained from polynomial fit results of
arriving time epochs of pulses. The local TOAs of the mean pulses
for individual observing sessions are determined from the maximum
cross correlation between the observed mean pulses and a Gaussian
profile template. The profile template is a mean pulse with high
signal-to-noise ratio, obtained by summing the best-quality mean
pulses over several observing sessions. Correction of TOAs to the
solar system barycenter can be done using
TEMPO2\footnote{http://www.atnf.csiro.au/research/pulsar/tempo2.}
program with the Jet Propulsion Laboratory DE405 ephemeris (Standish
1998). These TOAs are then weighted by the inverse squares of their
estimated uncertainty. Since the rotational period is nearly
constant, these observable quantities, $\nu$, $\dot{\nu}$ and
$\ddot{\nu}$ can be obtained by fitting the phases to the third
order of its Taylor expansion over a time span $t_{\rm s}$,
\begin{equation}\label{phase}
\Phi_i = {\Phi} + \nu (t_i-t) + \frac{1}{2}\dot \nu (t_i-t)^2 +
\frac{1}{6}\ddot\nu (t_i-t)^3.
\end{equation}

One can thus get the values of $\nu$, $\dot{\nu}$ and $\ddot{\nu}$
at $t$ from fitting to Equation~(1) for independent $N$ data blocks
around $t$, i.e. $i=1,...,N$. Apparently, these observational
quantities obtained this way are not instantaneous results, rather,
the ``averaged'' results over each data block (i.e. over each
$t_{\rm s}$) and extrapolated to $t$, which are not necessarily the
same as the instantaneous values (denoted as $\nu_{\rm I}$ and
$\dot{\nu}_ {\rm I}$). Thus, they are called ``averaged'' values
(denoted as $\nu_{\rm A}$ and $\dot{\nu}_{\rm A}$) in this work.
Usually, $t_{\rm s}$ is much less than pulsar's spin-down age
$\tau_c$, thus the differences between instantaneous values and
``averaged'' values are not significant, and consequently $\nu_{\rm
A}$ and $\dot{\nu}_{\rm A}$ are good approximations for $\nu_{\rm
I}$ and $\dot{\nu}_{\rm I}$ in most cases. However, it has been
found recently that oscillations of the ``apparent" magnetic fields
of neutron stars are responsible for the observed signs and
magnitudes of $\ddot{\nu}$, the second derivative of frequency, and
braking indices (Biryukov et al. 2012; Pons et al. 2012; Zhang \&
Xie 2012a, 2012b). We further suggested that the oscillation time
scales are between 10-100 yr, comparable to $t_{\rm s}$, thus making
the fitted spin-down parameters different from the true and
instantaneous spin-down parameters. Similarly, considerable biases
may also exist when fitting the glitch recovery data, since the
glitch recovery time scales are also comparable with $t_{\rm s}$.

In section 2, we simulate several pulsar timing data analysis
procedures for glitch recoveries, and find that the glitch
parameters, obtained from the averaged $\nu_{\rm A}$ and
$\dot{\nu}_{\rm A}$, have significant systematic biases compared
with that obtained with the instantaneous $\nu_{\rm I}$ and
$\dot{\nu}_{\rm I}$. In order to get the true glitch parameters with
the reported, yet averaged glitch recovery data $\nu_{\rm O}$ and
$\dot{\nu_{\rm O}}$, a phenomenological or physical glitch model is
needed to be combined with simulations. We thus present a
phenomenological spin-down model during a glitch recovery, and model
several slow glitch recovery events and the recovery of a giant
classical glitch in section 3. In section 4, we test four fitting
procedures based on the phenomenological spin-down model and find
that the best method is taking a very high order polynomial to fit
the phase and then taking its derivatives to obtain $\nu(t)$ and
$\dot\nu(t)$. In Section 5, we discuss how to obtain the model
parameters of glitch recoveries more accurately. The results are
summarized in section 6.

\section{Simulating Data Analysis of Glitch Recoveries}

\subsection{Simulation for $\dot\nu$-fitting procedure}

By fitting the TOA set $\{\Phi(t_i)\}$ to Equation (\ref{phase}),
one can get $\{\nu(t)\}$ and $\{\dot\nu(t)\}$. When $\{\nu(t)\}$ and
$\{\dot\nu(t)\}$ show exponential relaxations, their variations
following the jump at epoch $t_{\rm 0}$ can be described as the
following empirical functions (e.g. Yuan et al. 2010, Roy et al.
2012),
\begin{equation}\label{relaxation}
\nu (t) = {\nu _0}(t)+\Delta\nu_{\rm p} +\Delta\dot\nu_{\rm p}
\Delta t + \frac{1}{2}\Delta\ddot\nu_{\rm p}\Delta
t^2+\sum_j\Delta\nu _{{\rm d}j} e^{- \Delta t /\tau_j},
\end{equation}
and
\begin{equation}\label{relaxation2}
\dot\nu (t) = {\dot\nu _0}(t) +\Delta\dot\nu_{\rm p}
+\Delta\ddot\nu_{\rm p} \Delta t +\sum_j\Delta\dot\nu _{{\rm d}j}
e^{- \Delta t /\tau_j},
\end{equation}
where $\Delta t=t-t_{\rm 0}$, $\Delta\nu_{\rm p}$ and
$\Delta\dot\nu_{\rm p}$ are permanent changes in $\nu$ and $\dot\nu$
relative to the pre-glitch solution $\nu_{0}(t)$ and
$\dot\nu_{0}(t)$, $\Delta\nu_{{\rm d}j}$ is the amplitude of the
$j$th decaying component with a time constant $\tau_j$, and
$\Delta\dot\nu _{{\rm d}j}=-\Delta\nu_{{\rm d}j}/\tau_j$. One can
get the glitch parameters $\Delta\nu_{\rm p}$, $\Delta\dot\nu_{\rm
p}$, $\Delta\ddot\nu_{\rm p}$, $\Delta\nu_{{\rm d}j}$,
$\Delta\dot\nu_{{\rm d}j}$ and $\tau$ by fitting $\nu(t)$ and
$\dot\nu(t)$ to Equations~(\ref{relaxation}) and
(\ref{relaxation2}), respectively. The two functions describe the
post-glitch behaviors fairly well, especially for the case of a long
term recovery, and usually multiple decay terms with different decay
time constants can be fitted (e.g. there are up to five exponentials
are fitted for Vela 2000 and 2004 glitches; Dodson et al. 2002,
2007). For simplicity, the cases that $\nu$ varies as one
exponential decay term or two exponential decay terms are assumed in
the following simulations.

\textbf{Slow glitches are characterized
by a gradual increase in $\nu$ with a long time scale of several
months, accompanied by a rapid decrease in $|\dot\nu|$ by a few percent, which is sometimes even shorter than the observation interval and thus
cannot be seen. Then $|\dot\nu|$ experiences an
exponential increase back to its initial value with the same time
scale as that of $\nu$ increase (Shabanova 2005).} Analogous to the classical glitches, we suggest
that the slow glitches can be described by the following two
functions:
\begin{equation}\label{Srelaxation}
\nu (t) = {\nu _0}(t)+\Delta\nu_{\rm p} +\Delta\dot\nu_{\rm p}
\Delta t + \frac{1}{2}\Delta\ddot\nu_{\rm p}\Delta
t^2+\sum_j\Delta\nu _{{\rm d}j} (1-e^{- \Delta t /\tau_j}),
\end{equation}
and
\begin{equation}\label{Srelaxation2}
\dot\nu (t) = {\dot\nu _0}(t) +\Delta\dot\nu_{\rm p}
+\Delta\ddot\nu_{\rm p} \Delta t +\sum_j(-\Delta\dot\nu _{{\rm d}j})
e^{- \Delta t /\tau_j},
\end{equation}
where the parameters are the same as those in Equations
(\ref{relaxation}) and (\ref{relaxation2}).

\subsubsection{Simulation for One Decay Term }

Since the glitch or slow glitch recoveries can be described by
Equations (\ref{relaxation})-(\ref{Srelaxation2}), some simple
models can also be derived from them. For a classical glitch, we
simply assume
\begin{equation}\label{nu_t1}
 \nu(t)=\Delta\nu_{\rm d}\exp{(-\Delta t/\tau)},
\end{equation}
i.e., ${\nu_0}=\Delta\nu_{\rm p}=\Delta\dot\nu_{\rm
p}=\Delta\ddot\nu_{\rm p}=0$. We will use this equation to produce
simulated data, and obtain the ``instantaneous'' $\nu_{\rm
I}=\nu(t)$ and $\dot{\nu}_{\rm I}=d\nu(t)/dt$, with the parameters
($\Delta\nu_d$ and $\tau$) given later. On the other hand, the
``averaged'' values are obtained by the following procedure.
Firstly, we get the phase by $\Phi(t)=\int^{t}_{t_0}\nu(t')dt'$. For
convenience we take $t_0=0$. However, in practice $t_0$ cannot be
known precisely due to discontinuous observations; we will show
later that this will cause some uncertainty in estimating the
parameters of a classical glitch, but not so for slow glitches. We
assume a certain time interval $\Delta T_{\rm int}$ between each two
nearby TOAs, i.e. $\Delta T_{\rm int}\equiv t_{i+1}-t_{i}$ is a
constant. We set ten adjacent TOAs in one block (i.e. $N=10$ in
Equation~(\ref{phase})), and the latter five TOAs are used as the
first five TOAs in the next block. We then fit the TOA blocks to
Equation~(\ref{phase}) to obtain $\nu_{\rm A}$ and $\dot\nu_{\rm
A}$, which are the fitted coefficients of $\nu$ term and $\dot\nu$
term of the equation, respectively. The time $t$ for $\nu_{\rm A}$
and $\dot{\nu}_{\rm A}$ is taken as the middle epoch of each block,
i.e., $t=(t_5+t_6)/2$, and is also ``averaged'' (e.g. Yuan et al.
2010).

\begin{figure}
\begin{center}
\includegraphics[angle=0,scale=0.5]{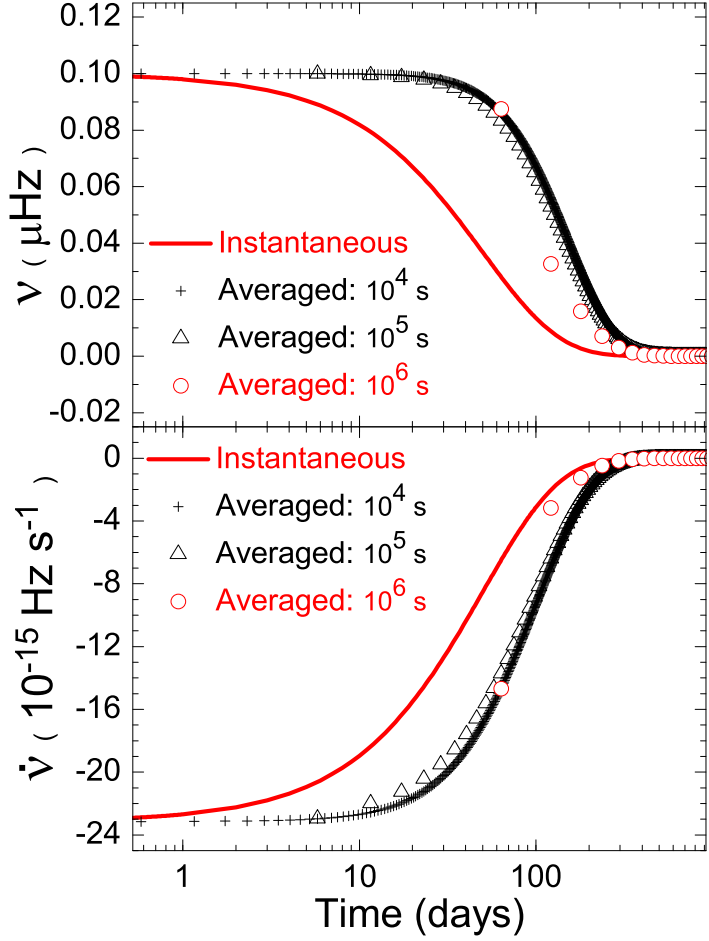}
\caption{Variations of $\nu$ (upper panel) and $\dot\nu$ (bottom
panel) for a simulated classical glitch with \emph{one decay term}.
Solid lines represent their instantaneous values; circles,
triangles, and crosses represent the averaged values obtained by
fitting the simulated TOAs to Equation~(\ref{phase}) for the cases
of $\Delta T_{\rm int}=10^{4}$, $10^{5}$, and $10^{6}$~s,
respectively.} \label{Fig:0}
\end{center}
\end{figure}

In Figure~\ref{Fig:0}, we show these instantaneous values and
averaged values with different $\Delta T_{\rm int}$ for a glitch
with $\Delta\nu_{\rm d}=0.1~\textmd{$\mu$Hz}$ and
$\tau=50~\textmd{days}$. One can see that both $\nu_{\rm A}$ and
$\dot{\nu}_{\rm A}$ have remarkably different decay profiles from
$\nu_{\rm I}$ and $\dot{\nu}_{\rm I}$ during the recovery process,
respectively. This systematic biases are independent of $\Delta
T_{\rm int}$, and it seems that the recovery time-scale $\tau$ is
the key parameter that is mainly biased. By fitting $\nu_{\rm A}$
and $\dot{\nu}_{\rm A}$ to Equation~(\ref{relaxation}) and
Equation~(\ref{relaxation2}), respectively, we find that all the
recovery time scales of $\nu_{\rm A}$ and $\dot{\nu}_{\rm A}$ are
much longer than the time scale of $50$~days (e.g. $\tau\approx 95$
day for $\Delta T_{\rm int}=10^4$ s). The systematic differences
between the decay profiles of $\nu_{\rm A}$ or $\dot{\nu}_{\rm A}$
and the profile of $\nu_{\rm I}$ or $\dot{\nu}_{\rm I}$ are
considerable, and apparently caused by the procedure of fitting TOAs
to Equation~(\ref{phase}); thus for higher order fits, one cannot
consider the first order coefficient to be the ``frequency". {\it
This procedure is thus abandoned for glitch data analysis in the
following.}

\begin{figure*}
\begin{center}
\includegraphics[angle=0,scale=0.6]{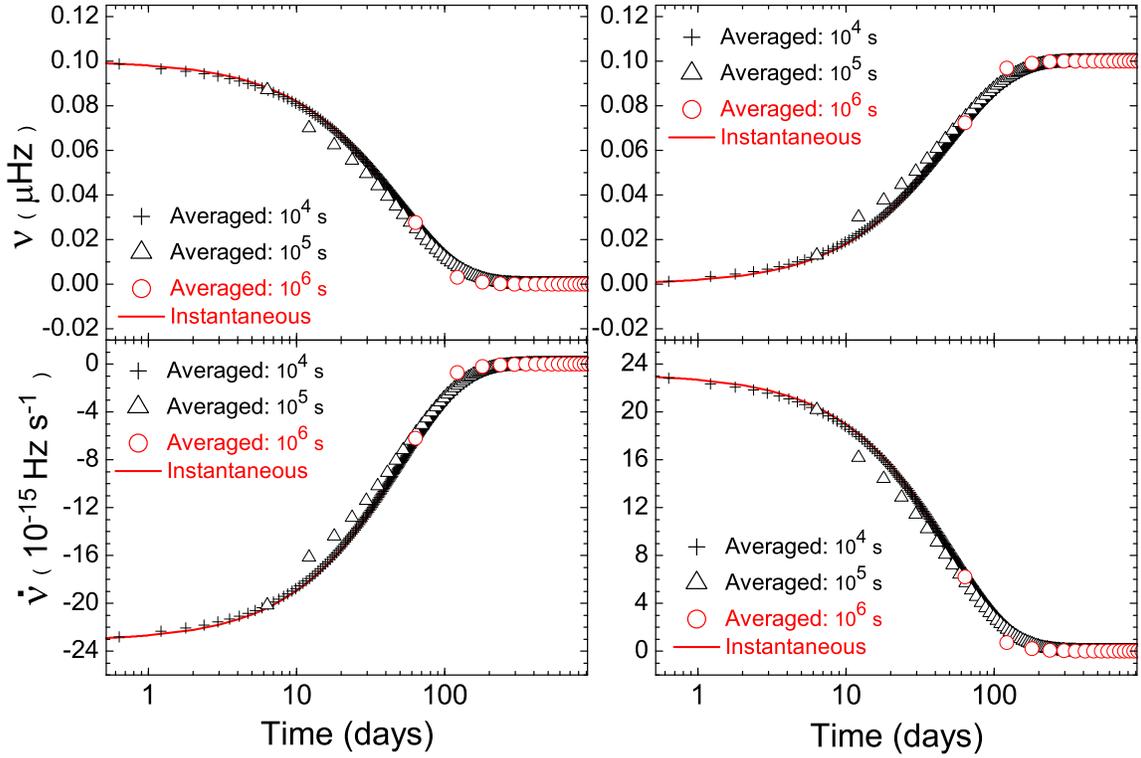}
\caption{Variations of $\nu$ and $\dot\nu$ for a simulated classical
glitch (left panels) and a simulated slow glitch (right panels) with
\emph{one decay term}. Solid lines represent their instantaneous
values; circles, triangles, and crosses represent the averaged
values obtained by fitting the simulated TOAs to
Equations~(\ref{phase2}) (to get $\nu_{\rm A}$) and (\ref{phase3})
(to get $\dot \nu_{\rm A}$) for the cases of $\Delta T_{\rm
int}=10^{4}$, $10^{5}$, and $10^{6}$~s, respectively.} \label{Fig:1}
\end{center}
\end{figure*}

However with the TEMPO2 software, $\nu$ may be obtained from the TOAs by fitting to
\begin{equation}\label{phase2}
\Phi_i = {\Phi} + \nu (t_i-t),
\end{equation}
and $\dot\nu$ may be obtained by fitting to
\begin{equation}\label{phase3}
\Phi_i = {\Phi} + \nu (t_i-t) + \frac{1}{2}\dot \nu (t_i-t)^2,
\end{equation}
i.e. the first two or three terms of Equations~(\ref{phase}),
respectively (Yu 2013). Here, we first fit the TOA blocks to
Equation~(\ref{phase2}) to obtain $\nu_{\rm A}$, which is the fitted
coefficients of $\nu$ term of Equation ~(\ref{phase2}). We then
separately fit the TOA blocks to Equation~(\ref{phase3}) to obtain
$\dot\nu_{\rm A}$, which is the fitted coefficients of $\dot\nu$
term of Equation ~(\ref{phase3}). In the left panels of
Figure~\ref{Fig:1}, we show the instantaneous and averaged values
obtained this way, with different $\Delta T_{\rm int}$ for a glitch
with the same $\Delta\nu_{\rm d}$ and $\tau$. Clearly now the
profiles of both $\nu_{\rm A}$ and $\dot{\nu}_{\rm A}$ follow that
of $\nu_{\rm I}$ and $\dot{\nu}_{\rm I}$ without obvious
distortions. By fitting to Equation~(\ref{relaxation}) or
Equation~(\ref{relaxation2}), we find that all the recovery time
scales of $\nu_{\rm A}$ or $\dot{\nu}_{\rm A}$ equal the time scale
of $50$~days, i.e. $\tau$ has not been biased.

We can then obtain the normally reported glitch parameters
$\Delta\nu_{\rm d}$ and $\Delta\dot\nu_{\rm d}$, as listed in Table
\ref{Tab:1}, by fitting $\{\nu_{\rm A}\}$ or $\{\dot{\nu}_{\rm A}\}$
to Equation~(\ref{relaxation}) or Equation~(\ref{relaxation2}) with
different $\Delta T_{\rm int}$; for comparison we also list
$\Delta\nu_{\rm d}$ and $\Delta\dot\nu_{\rm d}$ obtained from
$\nu_{\rm I}$ and $\dot{\nu}_{\rm I}$. One can see that ``averaged''
$\Delta\nu_{\rm d}$ or $\Delta\dot{\nu}_{\rm d}$ (denoted as
$\Delta\nu_{\rm dA}$ or $\Delta\dot{\nu}_{\rm dA}$ hereafter) have
systematic differences from instantaneous $\Delta\nu_{\rm d}$ or
$\Delta\dot{\nu}_{\rm d}$ (denoted as $\Delta\nu_{\rm dI}$ or
$\Delta\dot{\nu}_{\rm dI}$ hereafter). For $\Delta T_{\rm
int}=10^4~{\rm s}$, the differences are tiny and the glitch
parameters can be restored satisfactorily; however, for $\Delta
T_{\rm int}\geqslant 10^5~{\rm s}$, both the ``averaged''
$\Delta\nu_{\rm d}$ and $\Delta\dot{\nu}_{\rm d}$ may be
considerably smaller than the instantaneous $\Delta\nu_{\rm d}$ and
$\Delta\dot{\nu}_{\rm d}$, respectively.

For a slow glitch, we assume
\begin{equation}\label{nu_t2}
\nu(t)=\Delta\nu_{\rm d}(1-\exp{(-\Delta t/\tau)}),
\end{equation}
where $t_0=0$ and $\Delta\nu_{\rm d}=0.1~\textmd{$\mu$Hz}$ and
$\tau=50~\textmd{days}$. We show the averaged glitch parameters and
profiles with different $\Delta T_{\rm int}$ in Table \ref{Tab:1}
and the right panels of Figure~\ref{Fig:1}, as well as those
instantaneous ones. One can see that we always have $\Delta\nu_{\rm
dA}=\Delta\nu_{\rm dI}$ for any $\Delta T_{\rm int}$, since
$\Delta\nu_{\rm dA}$ \textbf{is determined by the differences of
$|\dot\nu_{\rm A}|$ between the data points slightly before the
starting point of the glitch and the data points at the end of the
recovery, and both of them are always available for slow
glitch observations.} However, $\Delta\dot{\nu}_{\rm dA}$ is biased in the same
way as for the simulated classical glitch.

\begin{deluxetable}{lccc|ccccccccccc}
\tabletypesize{\scriptsize} \tablecaption{$\Delta\nu_{\rm d}$
($\mu$Hz), $\Delta\dot\nu_{\rm d}$ ($10^{-15}$~Hz~s$^{-1}$) and
$\tau$ (days) for classical (upper part) and slow (bottom part)
glitch simulations. The data in left part and right part are results
from one decay term and two decay term simulations, respectively.
``Instantaneous" represents the instantaneous values, and the
different values of $\Delta T$ represent the time intervals between
each TOA for the ``averaged" values. }

\tablewidth{0pt}\tablehead{\colhead{$ $} & \colhead{$\Delta\nu_{\rm
d}$} & \colhead{$\Delta\dot\nu_{\rm d}$} & $\tau$ &
\colhead{$\Delta\nu_{\rm d}$} & \colhead{$\Delta\dot\nu_{\rm d}$} &
$\tau$}

\startdata
Instantaneous                         &0.100 & -23.15 & 50.00 & (0.19, 0.119) & (-102.8, -9.4)  & (21.4,~147.0)\\
$\Delta T_{\rm int}=10^{4}~{\rm s}$   &0.099 & -22.88 & 50.00 & (0.18, 0.119) & (-100.2, -9.4)  & (21.3,~145.8)\\
$\Delta T_{\rm int}=10^{5}~{\rm s}$   &0.089 & -20.67 & 50.00 & (0.13, 0.139) & (-100.4, -16.8) & (14.5,~95.6) \\
$\Delta T_{\rm int}=10^{6}~{\rm s}$   &0.041 & -9.53  & 49.52 & (2.73, 0.081) & (-2960.9, -6.4) & (10.7,~146.8)\\
\hline
Instantaneous                         &0.100 & 23.15 & 50.00 & (0.19, 0.119) & (102.8, 9.4)  & (21.4,~147.0)\\
$\Delta T_{\rm int}=10^{4}~{\rm s}$   &0.100 & 23.04 & 49.95 & (0.19, 0.120) & (100.2, 9.4)  & (21.3,~145.8)\\
$\Delta T_{\rm int}=10^{5}~{\rm s}$   &0.100 & 20.67 & 49.99 & (0.19, 0.122) & (100.4, 16.8)  & (14.5,~95.7) \\
$\Delta T_{\rm int}=10^{6}~{\rm s}$   &0.100 & 9.88  & 49.36 & (0.25, 0.055) & (2959.0, 6.4)  & (10.7,~146.8) \\
\hline

\enddata
\label{Tab:1}
\end{deluxetable}

\subsubsection{Simulation for Two Decay Terms }

\begin{figure*}
\begin{center}
\includegraphics[angle=0,scale=0.6]{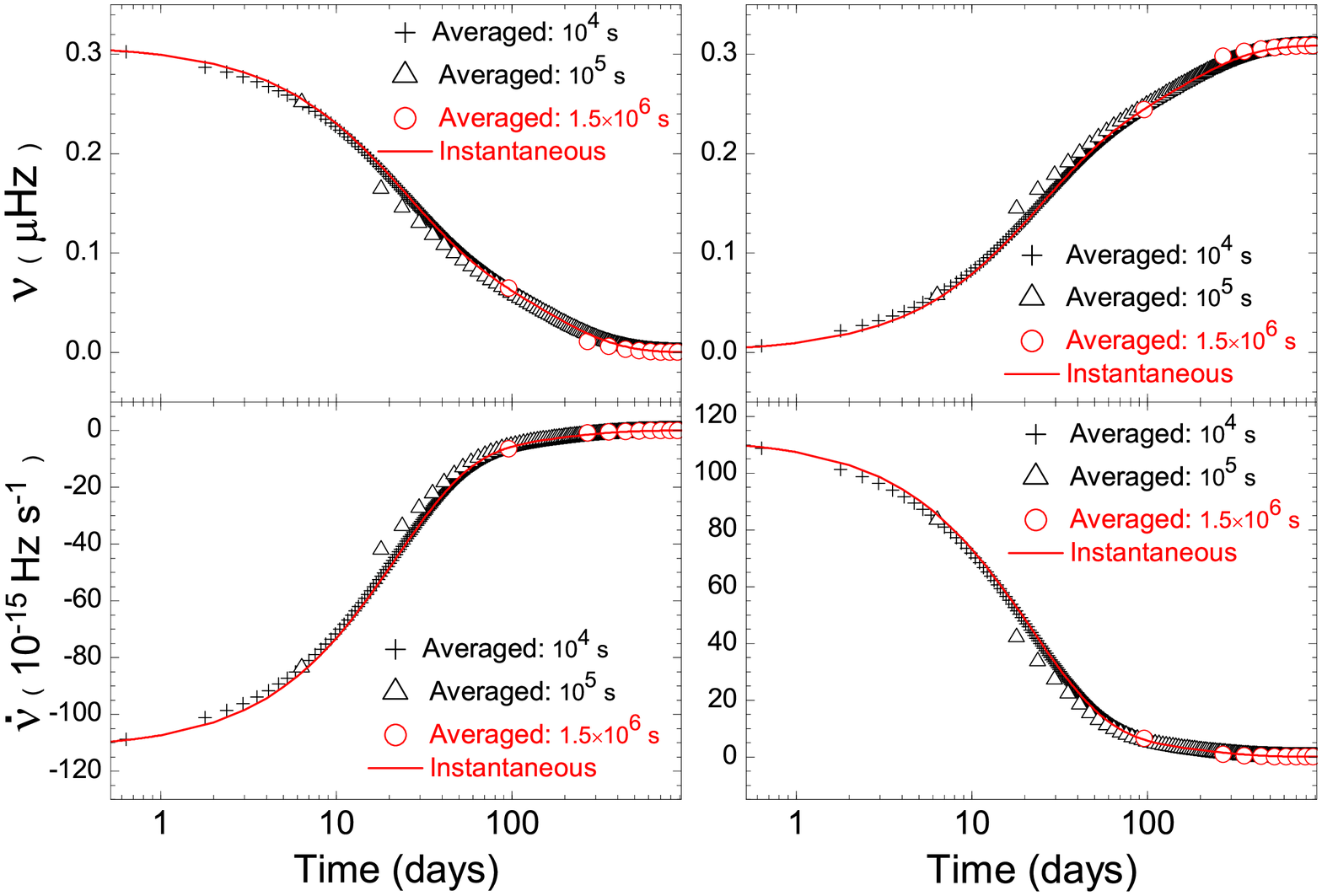}
\caption{Variations of $\nu$ and $\dot\nu$ for a simulated classical
glitch (left panels) and a simulated slow glitch (right panels) with
\emph{two decay terms}. Solid lines represent their instantaneous
values; circles, triangles, and crosses represent the averaged
values obtained by fitting the simulated TOAs to
Equations~(\ref{phase2}) (to get $\nu_{\rm A}$) and (\ref{phase3})
(to get $\dot \nu_{\rm A}$) for the cases of $\Delta T_{\rm
int}=1.5\times10^{6}$, $10^{5}$, and $10^{4}$~s, respectively.}
\label{Fig:2}

\end{center}
\end{figure*}

We simply assume $\nu(t) =\Delta\nu_{{\rm d}1}
\exp{(-t/\tau_1)}+\Delta\nu_{{\rm d}2} \exp{(-t/\tau_2)}$ for a
classical glitch with two decay terms, where $\Delta\nu_{{\rm
d}1}=0.19 $ $\mu$Hz, $\tau_1=21.4~\textmd{days}$ and
$\Delta\nu_{{\rm d}2}=0.119$ $\mu$Hz, $\tau_2=147~\textmd{days}$
(the parameters are adopted from pulsar B2334+61 for its very large
glitch between 2005 August 26 and September 8, Yuan et al. 2010). We
also assume $\nu(t) =\Delta\nu_{{\rm d}1}(1-
\exp{(-t/\tau_1)})+\Delta\nu_{{\rm d}2}(1-\exp{(-t/\tau_2)})$ for a
slow glitch with two decay terms. The instantaneous values and
averaged values are obtained with the same methods described above
and the main results are presented in Table \ref{Tab:1} and
Figure~\ref{Fig:2}, in which the similar results can be found with
the case of one decay term. For $\Delta T_{\rm int}=10^4~{\rm s}$,
the differences for all of $\tau$, $\Delta\nu_{\rm d}$ and
$\Delta\dot\nu_{\rm d}$ are tiny and the glitch parameters can be
restored satisfactorily. However, things for two decay terms are a
little more complicated. For $\Delta T_{\rm int}\gtrsim 10^5~{\rm
s}$, though the data points still converged to the instantaneous
values as shown in Figure \ref{Fig:2} (i.e. variation trends are the
same), the fitted glitch parameters (including $\tau$) for each
components are still somewhat biased, and it seems that larger
$\Delta T_{\rm int}$ corresponds to a smaller $\tau$ for short
time-scale component. The biases are probably due to the fact that
the data are too sparse for $\Delta T_{\rm int}$.  \textbf{Actually if
$\tau$ of the short term decay component is comparable to or
shorter than the interval between the observations, then $\tau$ of
this component would be difficult to determine and can only be set as the internal.} Similar
results can be found for a slow glitch, but $\sum\Delta\nu_{\rm
dI}=\sum\Delta\nu_{\rm dA}$ is always kept.

The above simulations unveil significant biases caused by the
averaging procedures (i.e. fitting to Equation~(\ref{phase2}) and
Equation~(\ref{phase3})) for $\nu$ and $\dot{\nu}$ during glitch
recoveries. Thus, $\nu_{\rm A}$ and $\dot\nu_{\rm A}$ obtained this
way and $\nu_{\rm O}$ and $\dot\nu_{\rm O}$ (the subscript ``o''
means observed values) reported in literature should not be used
directly to test physical models. It should be noted that, for
one-decay-term cases, the reported amplitudes of $\Delta{\nu}$ and
$\Delta\dot{\nu}$ of a classical glitch are usually underestimated;
the reported amplitude of $\Delta\dot{\nu}$ of a slow glitch is also
underestimated, but that $\Delta{\nu}$ is not. However, these biases
were never noticed in almost all previous theoretical works modeling
glitch recoveries, and $\{\dot\nu(t)\}$ are usually directly
modeled, e.g. the post-glitch fits for Vela pulsar, Crab pulsar and
PSR 0525+21 with vortex creep model (Alpar et al. 1984b; Alpar,
Nandkumar, \& Pines 1985; Alpar et al. 1993; Chau et al. 1993;
Alpar, et al. 1996; Larson \& Link 2002), and the two-component
hydrodynamic model for Vela (van Eysden \& Melatos 2010). In these
works, the observed data $\{\dot\nu_{\rm O}(t)\}$ (i.e.
$\{\dot\nu_{\rm A}(t)\}$) are shown in $\dot\nu$-$t$ diagram and
fitted directly by theoretical models.

\subsection{Simulation for Phase-fitting procedure}

In order to make optimum use of all available data (Shemar \& Lyne
1996), the pulse phase induced by a glitch is usually fitted to the
following equation, which can give $\tau$ and $\Delta\nu_d$ (e.g. Yu
et al. 2013):

\begin{equation}\label{phase4}
\phi_{\rm g} = {\Delta\phi} + \nu_{\rm p}\Delta t +
\frac{1}{2}\dot\nu_{\rm p}\Delta t^2 + [1-e^{-(\Delta
t/\tau_i)}]\Delta\nu_{{\rm d}i}\tau_i,
\end{equation}
where the $\nu_{\rm p}$ and $\dot\nu_{\rm p}$ are the permanent
increments in $\nu$ and $\dot\nu$, respectively. However, it is
difficult to get $\tau_i$ directly by fitting to Equation
(\ref{phase4}). Actually, TEMPO2 implements only a linear fitting
algorithm, and one thus needs to have a good initial estimate for
$\tau_i$, which is estimated from post-glitch $\dot\nu$ variation by
eye inspecting. Then the estimated value was introduced into
Equation (\ref{phase4}) fits. By increasing or decreasing $\tau_i$,
a best estimated $\tau_i$ can be eventually found via minimum
post-fit $\chi^2$ (Yu et al. 2013). \emph{This procedure is widely
used for classical glitches, but not applied to slow glitches.}

We simulate the fitting procedure of Equation (\ref{phase4}) as
described above, and find that both $\tau$ and $\Delta\nu_{\rm d}$
can be obtained with high precision for one-decay-term case, if a
good initial estimate for $\tau$ is taken, as shown in Table
\ref{Tab:3}. For two-decay-term case, we also assume $\nu(t)
=\Delta\nu_{{\rm d}1} \exp{(-t/\tau_1)}+\Delta\nu_{{\rm d}2}
\exp{(-t/\tau_2)}$, where $\Delta\nu_{{\rm d}1}=0.119$ $\mu$Hz,
$\tau_1=147~\textmd{days}$ and $\Delta\nu_{{\rm d}2}=0.19$ $\mu$Hz,
$\tau_2=21.4~\textmd{days}$, and get the phase by
$\Phi(t)=\int\nu(t)dt$. Firstly, we estimate $\tau_1$ for the long
term one, and get the best-fit $\tau_1$ by fitting $\{\Phi(t_i)\}$
to Equation (\ref{phase4}). Then we fix $\tau_1$ and get the
timescale of the short term $\tau_2$ the same way. This process is
widely adopted in the data analysis of glitch recovery with TEMPO2
software (Yu et al. 2003). However, we find that the glitch
parameters of the long decay term in the two term cases, i.e
$\tau_1$ and $\Delta\nu_{\rm d1}$ obtained by this way are already
biased, as shown in Table \ref{Tab:3}.

\begin{deluxetable}{lcc|cccccccc}
\tabletypesize{\scriptsize} \tablecaption{$\Delta\nu_{\rm d}$
($\mu$Hz), $\tau$ (days) for classical glitch simulations with
direct phase-fit procedure. The data in left part and right part are
the results from one decay term and two decay term simulations,
respectively. Subscript ``1" indicates the long decay term in two
term cases.}

\tablewidth{0pt}\tablehead{\colhead{$ $} & \colhead{$\Delta\nu_{\rm
d}$} & $\tau$ & \colhead{$\Delta\nu_{\rm d1}$} & $\tau_1$}

\startdata

Instantaneous                         &0.100 & 50.00 &  0.119 & 147.0\\
$\Delta T_{\rm int}=10^{4}~{\rm s}$   &0.109 & 50.00 &  0.150 & 128.0\\
$\Delta T_{\rm int}=10^{5}~{\rm s}$   &0.108 & 50.00 &  0.151 & 127.6\\
$\Delta T_{\rm int}=10^{6}~{\rm s}$   &0.106 & 50.00 &  0.158 & 124.0\\
\hline
\enddata
\label{Tab:3}
\end{deluxetable}

\begin{figure*}
\centering
\includegraphics[scale=0.45]{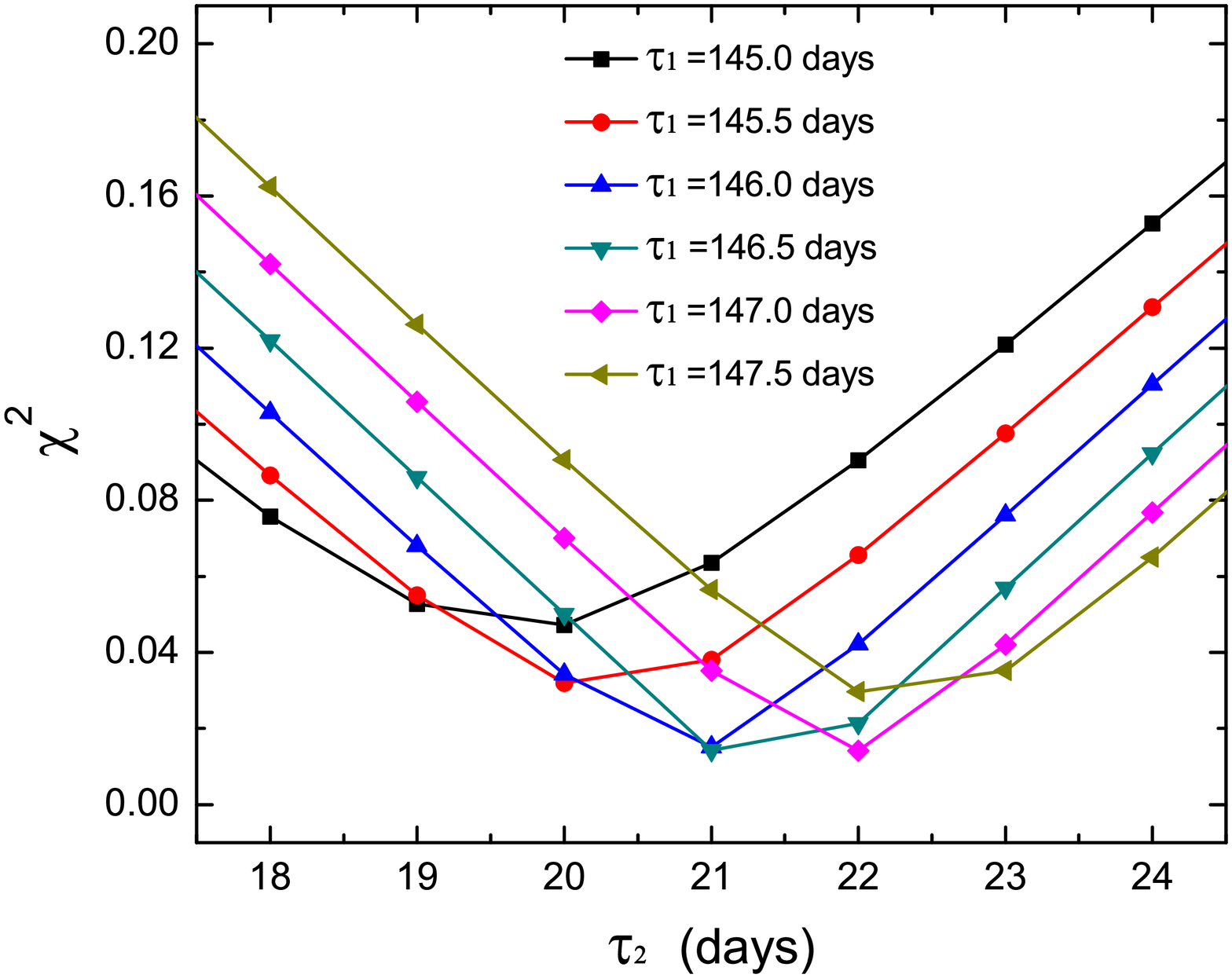}
\caption{The local minimum of $\chi^2$ produced by fitting
$\{\Phi(t_i)\}$ to Equation (\ref{phase4}). During each fitting,
$\tau_1$ is fixed to a certain value around 147 days.}
\label{Fig:10}
\end{figure*}

These biases are probably caused by the procedure that fitting the
long-decay-term and the short-decay-term in different steps; the
short-decay-term may slightly interfere the first fitting for
$\tau_1$ and $\Delta\nu_{{\rm d}1}$ of the long-decay-term, thus the
results are biased. If the biased $\tau_1$ is fixed, one will also
get a biased $\tau_2$ to fit $\{\Phi(t_i)\}$ again to Equation
(\ref{phase4}), since a local minimum $\chi^2$ will obtained, as
shown in Figure \ref{Fig:10}. Therefore, \emph{we suggest that the
two terms should be fitted simultaneously}.

If $\Delta T_{\rm int}\lesssim 10^4$ s, the simultaneous fits can be
realized by the following steps:

(i) Get $\{\dot\nu\}$ series by fitting $\{\Phi(t_i)\}$ to Equation
(\ref{phase3});

(ii) Estimate $\tau_1$ and $\tau_2$ by fitting $\{\dot\nu\}$ to
Equation (\ref{relaxation2}) (the calculation cost needed by this
fit is much lower than fitting to Equation (\ref{phase4}));

(iii) Use the estimated $\tau_1$ and $\tau_2$ as initial values and
fit $\{\Phi(t_i)\}$ again to Equation (\ref{phase4}), then the best
fitted $\tau_i$ and $\Delta\nu_{{\rm d}i}$ will be obtained.

The results of the simultaneous fit are $\Delta\nu_{{\rm d}1}=0.119$
$\mu$Hz, $\tau_1=147.0~\textmd{days}$ and $\Delta\nu_{{\rm
d}2}=0.190$ $\mu$Hz, $\tau_2=21.4~\textmd{days}$, which are exactly
the same with those introduced in the model; the results are
independent with $\Delta T_{\rm int}$. We also simulate the fitting
processes with different values of $\tau_i$ and $\Delta\nu_{{\rm
d}i}$, and all the glitch parameters are restored with relatively
small biases, some of which are even better than the previous
procedures for $\Delta T_{\rm int}\lesssim 10^4$ s. Here, we want
emphasize that the fitting procedures described in literature
are in chaos. Many authors adopted the procedure of fitting the TOAs
to Equation (\ref{phase4}) to obtain the pulsars parameters, and
using Equation (\ref{phase3}) to get $\{\dot\nu\}$, but only
Equation (\ref{phase}) is mentioned in their papers (Yu 2013). Thus,
we suggest that the exact fitting procedure should be described in
detail.

\section{A Phenomenological Spin-down Model}

In this section, we develop a phenomenological spin-down model to
describe the glitch and slow glitch recoveries, so that the model
can be a tool to simulate data to test the data analysis procedures
for the recoveries in the next section. Classically, a magnetic
dipole with a magnetic moment $M=BR^3$, rotating in vacuum with
angular velocity $\Omega$, emits electromagnetic radiation with a
total power $2M^2 \Omega^4/3c^3$. Assuming the pure magnetic dipole
radiation as the braking mechanism for a pulsar's spin-down, the
energy loss rate is then given by
\begin{equation}\label{dipole}
\dot{E}=I\Omega \dot \Omega =-\frac{2(BR^3\sin\chi)^2}{3c^3}{\Omega
^4},
\end{equation}
where $B$ is its dipole magnetic field at its magnetic pole, $R$ is its radius, $I$ is its moment of inertia.
Equation~(\ref{dipole}) is modified slightly in order to describe a glitch event,
\begin{equation}\label{redipole}
\dot\Omega\Omega^{-3} =-\frac{2(BR^3\sin\chi)^2}{3c^3I}G(t),
\end{equation}
in which $G(t)$ represents very small changes in the effective strength of dipole magnetic field $B\sin\chi$, or
the effective moment of inertia $I$ of both the pulsar and its magnetosphere during a glitch recovery. The left
hand are observable quantities, and the right hand are all theoretical quantities. In the following we assume
$G(t)=1+\kappa e^{-\Delta t/\tau}$. Then Equation~(\ref{redipole}) can be written as
\begin{equation}\label{rredipole}
\dot\nu\nu^{-3} =-H_0(1+\kappa e^{-\Delta t/\tau}),
\end{equation}
where $\dot\nu=\dot\Omega/2\pi$ and
$H_0=\frac{8\pi^2(BR^3\sin\chi)^2}{3c^3I}=1/2\tau_{\rm c}\nu_0^2$,
and $\tau_{\rm c}=-\nu/2\dot\nu$ is the characteristic age of a
pulsar.

Integrating and solving Equation~(\ref{rredipole}), we have
\begin{equation}\label{nu}
\nu(t)=\frac{\nu_0}{\sqrt{1+(\Delta t+\kappa\tau(1-e^{-(\Delta t)/\tau}))/\tau_{\rm c}}}.
\end{equation}
The derivative of $\nu$ is
\begin{equation}\label{dnu}
\dot\nu(t)=-\frac{(1+\kappa e^{-\Delta t/\tau})\nu_0/2\tau_{\rm
c}}{(1+(\Delta t+\kappa\tau(1-e^{-\Delta t/\tau}))/\tau_{\rm
c})^{3/2}}.
\end{equation}
We know $\Delta t\sim \tau \sim 100~{\rm days}$ and generally
$\tau_{\rm c}\gtrsim 10^4~{\rm years}$, and the term $|(\Delta
t+\kappa\tau(1-e^{-\Delta t/\tau}))/\tau_{\rm c}|\ll1$ and
$\kappa\ll 1$, the expression of $\nu$ and $\dot\nu$ can be
approximately written in the same forms of
Equations~(\ref{relaxation}) and (\ref{relaxation2}), which give
$\Delta\nu_{{\rm d}}=\nu_0\kappa\tau/2\tau_{\rm c}$ and
$\Delta\dot\nu_{{\rm d}}=-\nu_0\kappa/2\tau_{\rm c}$. Numerical
calculations show that Equations~(\ref{relaxation}) and
(\ref{relaxation2}) with these parameters give identical results as
Equations~(\ref{nu}) and (\ref{dnu}) for all known ranges of glitch
parameters. The expression of $\Delta\nu_{\rm p}$ and
$\Delta\dot\nu_{\rm p}$ that relate to the initial jumps of $\nu_0$
and $\dot\nu_0$, are not given by the model, since the glitch
relaxation processes are only considered here. It has been suggested
that these non-recoverable jumps are the consequence of permanent
dipole magnetic field increase during the glitch event (Lin and
Zhang 2004). $\Delta\ddot\nu_{\rm p}$ is the jump of timing
residual, which is beyond the scope of the present work. 

\textbf{In the following we attempt to apply this phenomenological model to fit the reported data of several slow glitches of B1822--09 and one classical glitch of B2334+61. Since the reported data points of $\nu$ and
$\dot\nu$ are too sparse (about one point per 150 days for B1822--09 or 50 days for B2334+61) and the TOAs of these glitches are not available in literature, we cannot apply our model to fit the reported to obtain both $\tau$ and $\kappa$ simultaneously, as that done in the above simulations. As a compromise, we focus only on determining $\kappa$ by applying our phenomenological model and simply take (the inevitably biased) $\tau$ obtained by fitting directly the reported $\nu$ and $\dot\nu$. Therefore $\kappa$ remains the only glitch recovery
parameter to be determined from observations in the following. Our main purpose here is to show the applicability of the our phenomenological model to describe glitch observations.}

\subsection{Modeling several slow glitches of B1822--09}

\begin{figure*}
\centering
\includegraphics[scale=0.75]{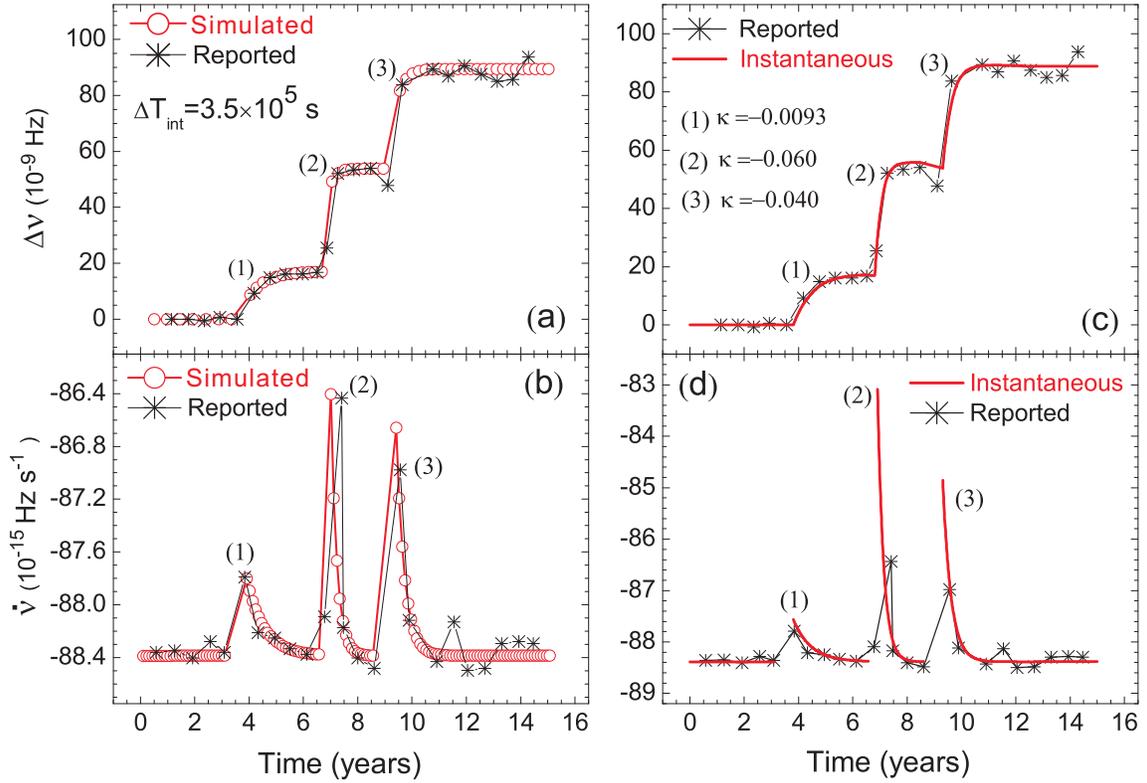}
\caption{Slow glitches of Pulsar B1822--09. Observational results are taken from Shabanova (2005). Upper panels:
variations of $\Delta\nu$ relative to the pre-glitch solution. Bottom panels: variations of $\dot{\nu}$. Left
panels: comparison between the reported and simulated (both are also time-averaged) $\Delta\nu$ and $\dot\nu$.
Right panels: comparison between the reported and restored (i.e. model-predicted) instantaneous $\Delta\nu$ and
$\dot\nu$.} \label{Fig:3}
\end{figure*}

We first model slow glitches because they are simpler than the
classical glitches, especially they have no jumps in $\nu$ and
$\dot\nu$, i.e. $\Delta\nu_{\rm p}=0$ and $\Delta\dot\nu_{\rm p}=0$.
Shabanova (2005) reported three slow glitches of B1822--09
(J1825-0935) over the 1995-2004 interval. The pulsar has
$\nu\simeq1.30~{\rm s}^{-1}$, a relatively large
$|\dot\nu|\simeq8.878\times10^{-14}~{\rm s}^{-2}$ (note
$\dot\nu<0$), implying $\tau_{\rm c}\simeq232$~kyr and
$B\simeq6.43\times10^{12}$~G. As shown in Figure {\ref{Fig:3}}, the
pulsar experienced three slow glitches from 1995 to 2005. A gradual
increase in $\nu$ is well modeled by an exponential function with
timescales of $235$, $80$ and $110$ days, respectively. For
$\dot\nu$, the fractional decreases of $|\dot\nu|$ (i.e. increases
of $\dot\nu$) are about $0.7$, $2.7$ and $1.7$ percent,
respectively. The subsequent increases of $|\dot\nu|$ (i.e.
decreases of $\dot\nu$) back to the previous values with the same
time scales are also well described by exponential functions. The
third slow glitch was separately detected by Zou et al. (2004).

Since the detailed data on $\Delta T_{\rm int}$ are not reported in
literature, we assume an uniform TOA distribution with $\Delta
T_{\rm int}=3.5\times10^5$~s. We take the following steps in
modeling the observed data for each slow glitch event:

(i) We get our model-predicted TOAs with $\Delta T_{\rm int}$ by
integrating Equations~(\ref{relaxation}) or (\ref{nu}), with a
$\kappa$ for each slow glitch event.

(ii) We simulate the data analysis process by fitting every block of
ten adjacent TOAs to Equations~(\ref{phase2}) or
Equations~(\ref{phase3}) to obtain one set of $\nu_{\rm A}$ and
$\dot{\nu}_{\rm A}$; and the latter five TOAs are also used in the
next TOA block.

(iii) The above simulated $\nu_{\rm A}$ and $\dot{\nu}_{\rm A}$ are
compared with the reported glitch profile $\nu_{\rm O}$ and
$\dot{\nu}_{\rm O}$; $\kappa$ is adjusted until reasonable
agreements between them are reached.

With the above steps, we confirm that the slow glitch behavior can
be explained by our phenomenological model with $\kappa<0$. Our
modeling results are shown in Figure~\ref{Fig:3}. The fit parameter
$\kappa$ is $-0.0093$, $-0.06$ and $-0.04$ for the three slow glitch
events, respectively. In Table \ref{Tab:2} we show the relative
magnitudes of $\Delta\nu$ and $\Delta\dot\nu$ for the three slow
glitches; for comparison we also list in Table \ref{Tab:2} the
results for the giant classical glitch from B2334+61 obtained in the
next section. It is found that the relative magnitudes of
$\Delta\nu_{\rm A}$, $\Delta\nu_{\rm O}$ and $\Delta\nu_{\rm I}$ are
identical, i.e. $\Delta\nu_{\rm I}=\Delta\nu_{\rm A}=\Delta\nu_{\rm
O}$, as expected from the above simulations. It is also clear that
the instantaneous values of $\Delta\dot\nu$, \textbf{which are
calculated directly from the model with the parameters determined above,} are much
larger than the reported results in literature, e.g. the
$\Delta\dot\nu_{\rm I}$ are larger than two times of
$\Delta\dot\nu_{\rm O}$ for the second and third slow glitches.

\begin{deluxetable}{lcccccccccc}
\tabletypesize{\scriptsize} \tablecaption{The relative values of
$\Delta\nu$ and $\Delta\dot\nu$ for slow glitches of B1822--09 and
the giant classical glitch from B2334+61. In the classical glitch
part, the superscripts `${\rm i}$' and `${\rm ii}$' represent for
results of one-term fit and two-term fit, respectively.}
\tablewidth{0pt}

\tablehead{\colhead{} & \multicolumn{6}{c}{Slow Glitches of
B1822--09} & \multicolumn{4}{c}{Classical Glitch of B2334+61}\\

\cline{8-11}& \cline{1-6}

\colhead{$ $} & \colhead{$\Delta\nu_1/\nu_0$}&
\colhead{$\Delta\dot\nu_1/\dot\nu_0$}&
\colhead{$\Delta\nu_2/\nu_0$}&
\colhead{$\Delta\dot\nu_2/\dot\nu_0$}&
\colhead{$\Delta\nu_3/\nu_0$}&
\colhead{$\Delta\dot\nu_3/\dot\nu_0$}& \colhead{$\Delta\nu^{\rm
i}/\nu_0$}& \colhead{$\Delta\dot\nu^{\rm i}/\dot\nu_0$}&
\colhead{$\Delta\nu^{\rm ii}/\nu_0$}&
\colhead{$\Delta\dot\nu^{\rm ii}/\dot\nu_0$}\\

&($10^{-9}$)&(\%)&($10^{-9}$)& (\%) &($10^{-9}$)& (\%) &($10^{-9}$)&
(\%) &($10^{-9}$)& (\%)}

\startdata
Reported      & 12.9&0.7 & 28.6 & 2.7 & 25.2 & 1.7 &75.8& -2.96 &75.8& -2.96\\
Simulated      &13.2 &0.7 & 29.7 & 2.7 & 25.4 & 2.0 &35.6& -3.15 &54.5& -2.87\\
Instantaneous &13.2 &0.94& 29.7 & 6.0 & 25.4 & 4.0 &64.4& -3.85 &79.8& -3.98\\

\enddata
\label{Tab:2}
\end{deluxetable}

\subsection{Modeling one classical glitch of B2334+61}

\begin{figure*}
\centering
\includegraphics[angle=0,scale=0.75]{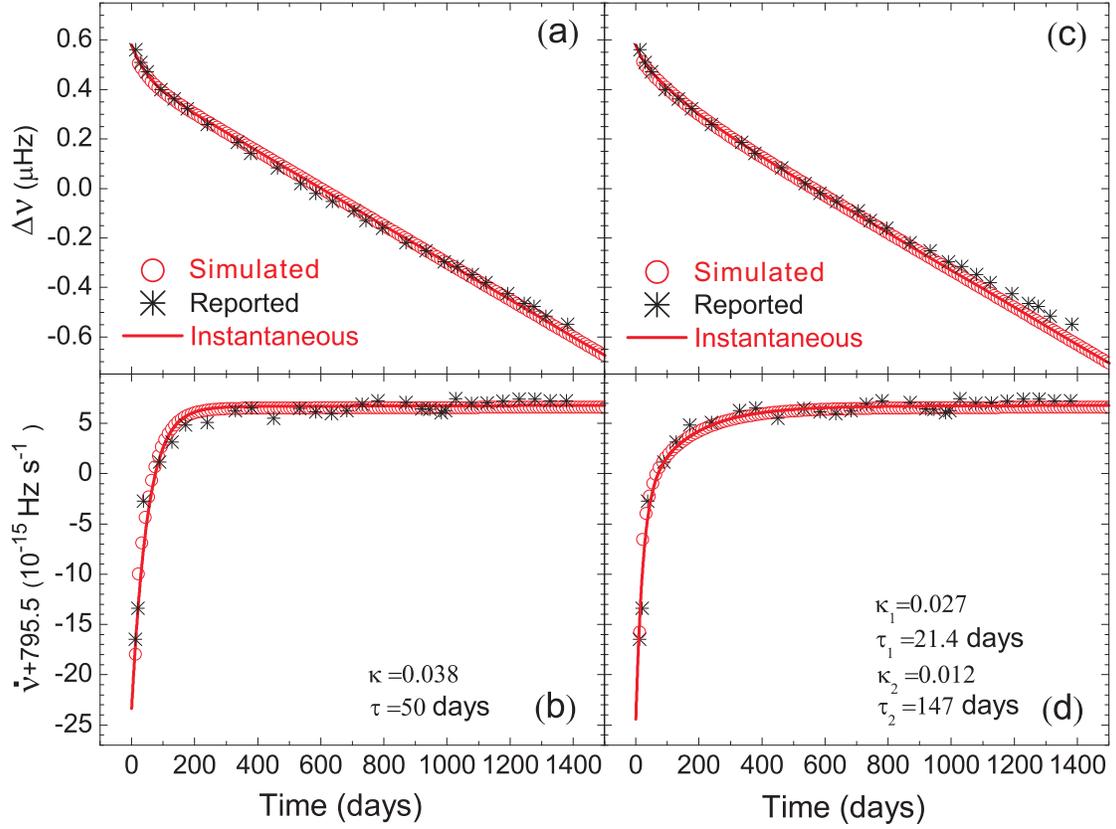}
\caption{The giant glitch of Pulsar B2334+61. Observational results
are taken from Yuan et al. 2010. $\Delta T_{\rm
int}=1.8\times10^5~{\rm s}$ is adopted in the fit. Upper panels:
variations of $\Delta\nu$. Bottom panels: variations of $\dot{\nu}$.
The left and right panels represent for models with one and two
decay components, respectively.} \label{Fig:4}
\end{figure*}

The pulsar PSR B2334+61 (PSR J2337+6151) was discovered in the
Princeton-NRAO survey using the 92 m radio telescope at Green Bank
in 1985 (Dewey et al. 1985). It has $\nu\simeq2.019~{\rm s^{-1}}$,
$\dot\nu\simeq-788.332\times10^{-15}~{\rm s^{-2}}$, $\tau_{\rm
c}\simeq4.1\times10^4~{\rm yr}$, and $B\simeq9.91\times10^{12}$~G.
It is located very close to the center of the supernova remnant
G114.3+0.3. Yuan et al. (2010) reported the timing observations of
PSR B2334+61 for seven years with the Nanshan 25~m telescope at
Urumqi Observatory. A very large glitch occurred between 2005 August
26 and September 8 (MJDs 53608 and 53621), the largest known glitch
ever observed, with a fractional frequency increase of
$\Delta\nu/\nu\sim20.5\times10^{-6}$. Yuan et al. (2010) obtained
each $\nu$, $\dot{\nu}$ and $\ddot{\nu}$ by fitting ten adjacent
TOAs to Equation~(\ref{phase}), and the latter five TOAs had also
been used as the first five TOAs in the next fit. The rotational
behavior during this glitch event is shown in Figure~\ref{Fig:4}. A
large jump of rotational frequency could be seen in the top panel
with $\Delta\nu\approx41\times10^{-6}$~Hz. The bottom panel shows a
very significant long-term increase in $|\dot{\nu}|$ after the time
of jump, and the corresponding braking indices are $10.5\pm0.2$ and
$46.8\pm0.3$ before and after the glitch, respectively. The recovery
process following the glitch was described by a dominant rapid
exponential decay with a time scale of $\sim21.4$~days and an
additional slower decay with a time scale of $\sim147$~days (Yuan et
al. 2010).

\begin{figure*}
\centering
\includegraphics[angle=0,scale=0.75]{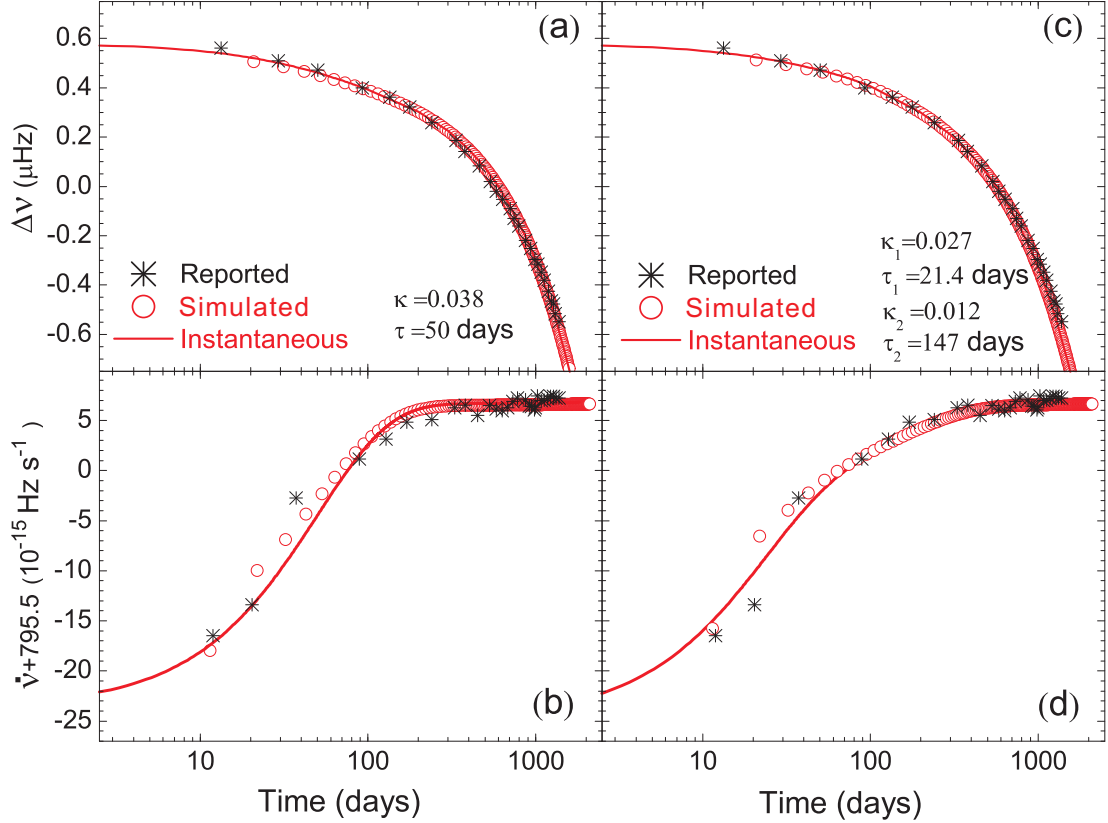}
\caption{The giant glitch of Pulsar B2334+61. Observational results
are taken from Yuan et al. 2010. $\Delta T_{\rm
int}=1.8\times10^5~{\rm s}$ is adopted in the fit. Upper panels:
variations of $\Delta\nu$. Bottom panels: variations of $\dot{\nu}$.
The left and right panels represent for models with one and two
decay components, respectively. This figure is the same with Fig.
\ref{Fig:4} except for a logarithmic abscissa.} \label{Fig:5}
\end{figure*}

\begin{figure*}
\centering
\includegraphics[angle=0,scale=0.7]{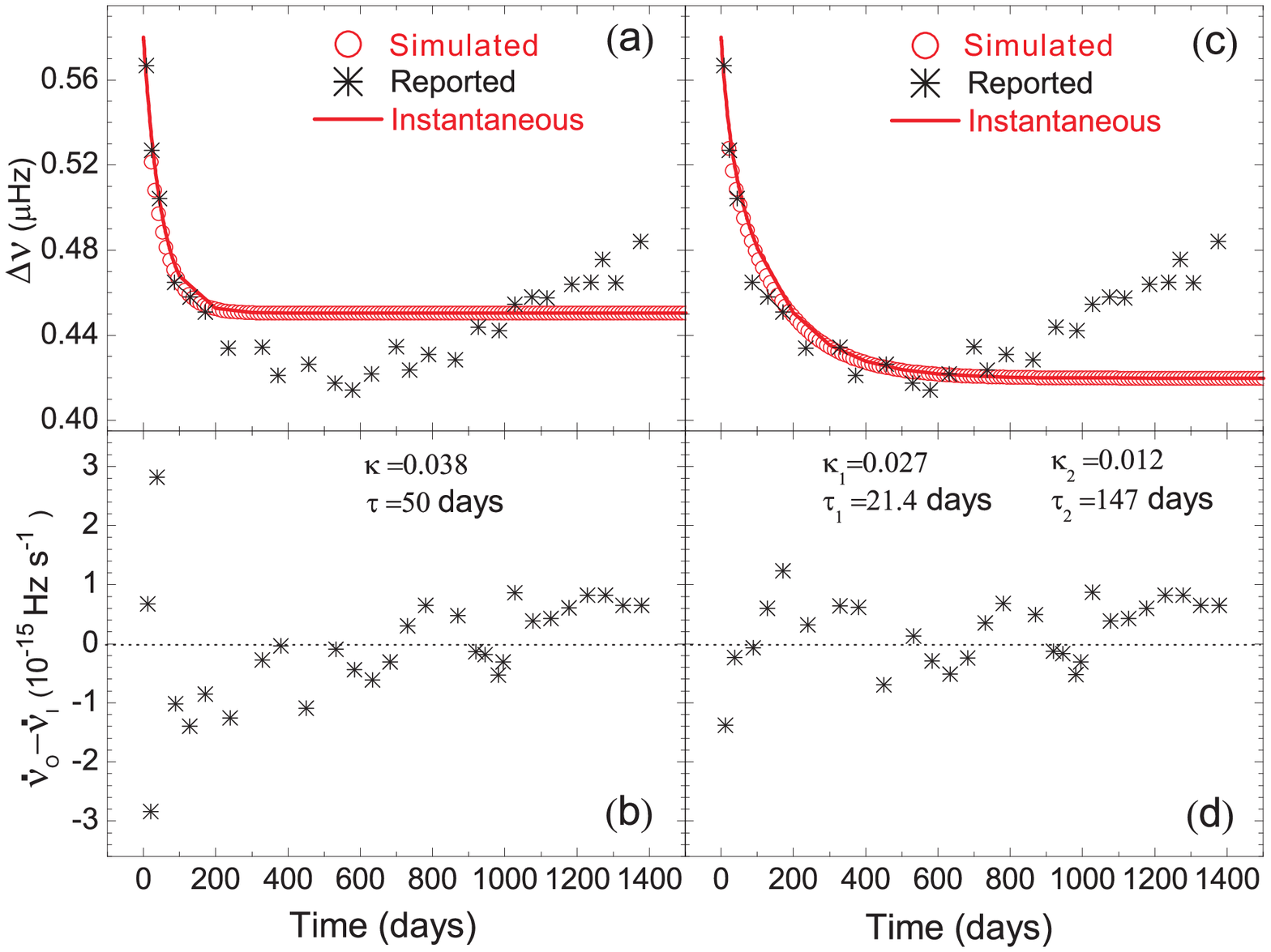}
\caption{The giant glitch of Pulsar B2334+61. Upper panels:
variations of $\Delta\nu$ from which the slope $\Delta\dot\nu_{\rm
p}$ is removed. Bottom panels: residuals of $\dot{\nu}_{\rm
O}-\dot\nu_{\rm I}$. The left and right panels represent for models
with one and two decay components, respectively.} \label{Fig:6}
\end{figure*}

We follow almost the same steps as for the slow glitches above to
model the reported, yet time-averaged glitch recovery data of this
classical glitch, with $\Delta T_{\rm int}=1.8\times10^5$~s; the
only difference is that a slope of $\Delta\dot\nu_{\rm
p}=-8.684\times10^{-15}~{\rm s}^{-2}$ is taken in
Equation~(\ref{relaxation}), following Lyne et al. (2000).

In the left panels of Figure~\ref{Fig:4}, we show the fits with one
exponential term $G(t)=(1+\kappa\exp{(-\Delta t/\tau)})$ \textbf{for
a comparison with the ``realistic'' simulation of two terms below}.
The best parameters for this glitch event are $\kappa_1=0.038$ and
$\tau=50$~days. We then model the glitch recovery process with
$G(t)=(1+\kappa_1\exp{(-\Delta t/\tau_1)}+\kappa_2\exp{(-\Delta
t/\tau_2)})$, as shown in the right panels of Figure~\ref{Fig:4}.
The best parameters for this glitch event are $\kappa_1=0.027$ and
$\kappa_2=0.012$ ($\tau_1=21.4$~days and $\tau_2=147$~days are fixed
by the observed values). Table \ref{Tab:2} gives the relative
magnitudes of $\Delta\nu$ and $\Delta\dot\nu$ for both fits. In
order to distinguish between the two fits, we show them in the
logarithmic coordinates in Figure~\ref{Fig:5}. Clearly the simulated
profiles of the two term fit match the reported ones better than
that of the one term fit. One can see that $|\Delta \dot\nu_{\rm
I}|$ are also slightly larger than the reported $|\Delta
\dot\nu_{\rm O}|$ for both the one-term fit and two-term fit.

In Figure~\ref{Fig:6} we show $\Delta\nu$ with the slope of
$\Delta\dot\nu_{\rm p}$ removed, and $\dot\nu_{\rm O}-\dot\nu_{\rm
I}$. It is clearly shown that one exponential term cannot fit the
observed data at the end of decay profile, and this is also the
reason why $\Delta \nu_{\rm I}$ is smaller than $\Delta \nu_{\rm O}$
for this fit, as given in Table \ref{Tab:2}. Thus, \textbf{the
one-term decay is ruled out}, and we focus on the two-term fit
below. Using $\nu_0$, $\tau_c$ and the determined $\kappa_1$ and
$\kappa_2$, we obtain these glitch parameters: $\Delta\nu_{\rm
d1}=0.039~{\rm \mu Hz}$, $\Delta\nu_{\rm d2}=0.119~{\rm \mu Hz}$,
$\Delta\dot\nu_{\rm d1}=-2.1\times10^{-14}~{\rm s}^{-2}$,
$\Delta\dot\nu_{\rm d2}=-9.38\times10^{-15}~{\rm s}^{-2}$; some of
them have significant differences from the reported results of Yuan
et al. (2010), in which $\Delta\nu_{\rm d1}=0.19~{\rm \mu Hz}$,
$\Delta\nu_{\rm d2}=0.119~{\rm \mu Hz}$, $\Delta\dot\nu_{\rm
d1}=-1.03\times10^{-13}~{\rm s}^{-2}$, $\Delta\dot\nu_{\rm
d2}=-9.37\times10^{-15}~{\rm s}^{-2}$.

In Figure~\ref{Fig:6}, one can also see an exponential increase of
$\nu$ after the glitch recovery, which is a very common, but not
well understood behavior (Lyne 1992, see an example of a Crab
glitch). We suggest that the exponential increase component is
probably a slow glitch, and the fact that a slow glitch following a
classical glitch recovery may be an important clue to the enigmas of
glitch phenomena.

\section{Testing Several Fitting Procedures Based on the Phenomenological Spin-down Model}

In section 3, we showed that the recovery processes of glitches and
slow glitches can be well modeled by the phenomenological model,
which can also be used to simulate a real glitch recoveries, in
order to fully test different fitting procedures. We get $\Phi(t)$
by integrating Equation (\ref{rredipole}) for a certain $\tau_i$ and
$\kappa_i$, and get the TOA set $\{\Phi(t_i)\}$ by assuming a
certain $\Delta T_{\rm int}$. We test the biases produced by the
following four fitting procedures in the section; \textbf{three of them
are discussed in section 2 for the simplified model of classical
and slow glitches. Here all four procedures are examined with a
more ``realistic'' model, i.e., our phenomenological spin-down model.}

\emph{Fitting Procedure I:} obtain $\{\dot\nu(t_i)\}$ by fitting
$\{\Phi(t_i)\}$ to Equation (\ref{phase}), and get $\tau$ and
$\Delta\nu_{\rm d}$ by fitting $\{\dot\nu(t_i)\}$ to Equation
(\ref{relaxation2}) (e.g. Roy et al. 2012, Espinoza et al. 2011,
Yuan et al. 2010, Zou et al. 2008, Zou et al. 2004, Dall'Osso et al.
2003, Urama 2002, Dodson et al. 2002, McCulloch et al. 1990). We
take $\tau=50$ days, $\kappa=0.03$ in the model, and show
instantaneous results and the fitted results for different $\Delta
T_{\rm int}$ in Table \ref{Tab:4}. The instantaneous $\Delta\nu_{\rm
d}$ is given by $\Delta\nu_{{\rm d}}=\nu_0\kappa\tau/2\tau_{\rm c}$.
It is found that both $\tau$ and $\Delta\nu_{\rm d}$ are seriously
biased. It is also noticed that the total time span $T_{\rm s}$
(i.e. the time span from the begin of the glitch to the end of the
recovery assumed), has considerable impact on the fitting results
else. We take higher order polynomials to fit the phase and find
that $\tau$ is also seriously biased, even worse than the lower
order one. Thus, if one takes a higher order polynomial and calls
the first (linear) term ``frequency" then this is clearly not a good
approximation, since a higher order polynomial would lead to this
not being the ``frequency", given that part of it is reabsorbed into
other coefficients.

\begin{deluxetable}{lcccccccccccccccc}
\tabletypesize{\scriptsize} \tablecaption{Classical glitch
simulations with the phenomenological model. $\Delta\nu_{\rm d}$
(0.1~$\mu$Hz) and $\tau$ (days) obtained by fitting procedure I.}

\tablewidth{0pt}\tablehead{ & \multicolumn{2}{c}{$T_{\rm s}=1~{\rm
year}$}  &  \multicolumn{2}{c}{$T_{\rm s}=3~{\rm years}$} &
 \multicolumn{2}{c}{$T_{\rm s}=5~{\rm
years}$} \\

& \colhead{$\tau$} & \colhead{$\Delta\nu_{\rm d} $} &
\colhead{$\tau$} & \colhead{$\Delta\nu_{\rm d} $} & \colhead{$\tau$}
& \colhead{$\Delta\nu_{\rm d} $} }

\startdata

Instantaneous                         & 50   & 1.01 &  50   & 1.01 &  50   & 1.01 \\
$\Delta T_{\rm int}=10^{4}~{\rm s}$   &213.38& 1.15 & 96.45 & 2.35 & 92.34 & 2.22 \\
$\Delta T_{\rm int}=10^{5}~{\rm s}$   &161.99& 6.49 & 90.74 & 2.51 & 87.52 & 2.11 \\
$\Delta T_{\rm int}=10^{6}~{\rm s}$   &27.02 & 3.68 & 37.75 & 3.07 & 38.75 & 3.04 \\
\hline
\enddata
\label{Tab:4}
\end{deluxetable}

\emph{Fitting Procedure II:} obtain $\{\dot\nu(t_i)\}$ by fitting
$\{\Phi(t_i)\}$ to Equation (\ref{phase3}), and get $\tau$ and
$\Delta\nu_{\rm d}$ by fitting $\{\dot\nu(t_i)\}$ to Equation
(\ref{relaxation2}) (e.g. Shabanova 2005, Shabanova 1998). Firstly,
we conduct the one decay component case and take $\tau=50$ days,
$\kappa=0.03$ in the model. The main results are shown in Table
\ref{Tab:5}. One can see that the instantaneous values can be well
restored for $\Delta T_{\rm int}=10^{4}~{\rm s}$, and the results
with $\Delta T_{\rm int}=10^{5}~{\rm s}$ are also good
approximations. Then, we conduct the two-component case and take
$\tau_1=21.7$ days, $\tau_2=147$ days, and $\kappa_1=0.131$,
$\kappa_2=0.012$ in the model. One can see that the instantaneous
values can only be restored for $\Delta T_{\rm int}=10^{4}~{\rm s}$.
It is noticed that $T_{\rm s}$ should be long enough for both the
one-component and two-component cases. Thus, this procedure is a
good approximation on very small $\Delta T_{\rm int}$.

\begin{deluxetable}{lcccccccccccccccc}
\tabletypesize{\scriptsize} \tablecaption{Classical glitch
simulations with the phenomenological model. $\Delta\nu_{\rm d}$
(0.1~$\mu$Hz) and $\tau$ (days) obtained by fitting procedure II.}

\tablewidth{0pt}\tablehead{ & \multicolumn{2}{c}{$T_{\rm s}=1~{\rm
year}$}  &  \multicolumn{2}{c}{$T_{\rm s}=3~{\rm years}$} &
 \multicolumn{2}{c}{$T_{\rm s}=5~{\rm
years}$} \\

& \colhead{$\tau$} & \colhead{$\Delta\nu_{\rm d} $} &
\colhead{$\tau$} & \colhead{$\Delta\nu_{\rm d} $} & \colhead{$\tau$}
& \colhead{$\Delta\nu_{\rm d} $} }

\startdata

Instantaneous                         & 50   & 1.01 &  50   & 1.01 &  50   & 1.01\\
$\Delta T_{\rm int}=10^{4}~{\rm s}$   &49.95 & 1.00 & 49.97 & 1.00 & 49.98 & 1.00 \\
$\Delta T_{\rm int}=10^{5}~{\rm s}$   &45.84 & 0.95 & 47.82 & 0.96 & 48.01 & 0.96\\
$\Delta T_{\rm int}=10^{6}~{\rm s}$   &22.23 & 5.02 & 27.37 & 3.15 & 27.85 & 2.42\\
\hline
Instantaneous                         & (21.40, 147.00) & (1.90, 1.19) & (21.40, 147.00) & (1.90, 1.19) & (21.40, 147.00) & (1.90, 1.19)\\
$\Delta T_{\rm int}=10^{4}~{\rm s}$   & (21.14, 113.58) & (1.85, 0.86) & (21.26, 144.47) & (1.87, 1.18) & (21.27, 145.45) & (2.33, 1.20)\\
$\Delta T_{\rm int}=10^{5}~{\rm s}$   & (2.26,   23.86) & (0.59, 1.70) & (9.58, 42.54)   & (0.76, 1.57) & (14.40, 81.41) & (1.90, 1.45)\\
\enddata
\label{Tab:5}
\end{deluxetable}

\emph{Fitting Procedure III:} get $\tau$ and $\Delta\nu_{\rm d}$
directly by fitting $\{\Phi(t_i)\}$ to Equation (\ref{phase4}) (Yu
et al. 2013, Edwards et al. 2006, Shemar \& Lyne 1996). Firstly, we
also conduct the one decay component case and take $\tau=50$ days,
$\kappa=0.03$ in the model. The main results are shown in upper part
of Table \ref{Tab:6}. It is found that the instantaneous values can
be well restored and the fit results is nearly independent of
$\Delta\ T_{\rm int}$. Then, we conduct the two-component case and
take $\tau_1=21.7$ days, $\tau_2=147$ days, and $\kappa_1=0.131$,
$\kappa_2=0.012$ in the model. We \emph{fit the two decay terms
simultaneously} (at a high computing cost) and the main results are
shown in the bottom part of Table \ref{Tab:6}. It is also found that
the instantaneous values can be restored satisfactorily and the
results are independent of $\Delta\ T_{\rm int}$. However, $T_{\rm
s}$ should not be too long for both the one component and two
components cases, which is opposite to procedure II.

\begin{deluxetable}{lcccccccccccccccc}
\tabletypesize{\scriptsize} \tablecaption{Classical glitch
simulations with the phenomenological model. $\Delta\nu_{\rm d}$
(0.1~$\mu$Hz) and $\tau$ (days) obtained by fitting procedure III.}

\tablewidth{0pt}\tablehead{ & \multicolumn{2}{c}{$T_{\rm s}=1~{\rm
year}$}  &  \multicolumn{2}{c}{$T_{\rm s}=3~{\rm years}$} &
 \multicolumn{2}{c}{$T_{\rm s}=5~{\rm
years}$} \\

& \colhead{$\tau$} & \colhead{$\Delta\nu_{\rm d} $} &
\colhead{$\tau$} & \colhead{$\Delta\nu_{\rm d} $} & \colhead{$\tau$}
& \colhead{$\Delta\nu_{\rm d} $} }

\startdata

Instantaneous                         &50.00 & 1.01 &  50   & 1.01 &  50   & 1.01\\
$\Delta T_{\rm int}=10^{4}~{\rm s}$   &50.16 & 1.02 & 53.13 & 0.99 & 67.23 & 0.83 \\
$\Delta T_{\rm int}=10^{5}~{\rm s}$   &50.16 & 1.02 & 53.11 & 0.99 & 67.01 & 0.84\\
$\Delta T_{\rm int}=10^{6}~{\rm s}$   &50.15 & 1.02 & 52.86 & 1.00 & 65.29 & 0.87\\
\hline
Instantaneous                         & (21.40, 147.00) & (1.90, 1.19) & (21.40, 147.00) & (1.90, 1.19) & (21.40, 147.00) & (1.90, 1.19)\\
$\Delta T_{\rm int}=10^{4}~{\rm s}$   & (21.38, 147.92) & (1.92, 1.20) & (21.73, 152.13) & (1.92, 1.20) & (18.98, 160.90) & (2.33, 1.20)\\
$\Delta T_{\rm int}=10^{5}~{\rm s}$   & (21.40, 147.93) & (1.92, 1.20) & (21.76, 152.17) & (1.92, 1.20) & (19.04, 160.87) & (2.31, 1.20)\\
$\Delta T_{\rm int}=10^{6}~{\rm s}$   & (21.40, 147.96) & (1.92, 1.20) & (21.77, 152.19) & (1.93, 1.20) & (19.35, 160.57) & (2.20, 1.20)\\
\enddata
\label{Tab:6}
\end{deluxetable}

\emph{Fitting Procedure IV:} the phase $\{\Phi(t_i)\}$ is fitted by
a very high order polynomial, such as
\begin{equation}\label{phase5}
\Phi(t) = {\Phi_0} + \nu (t-t_i) + \frac{1}{2}\dot \nu (t-t_i)^2 +
\frac{1}{6}\ddot\nu (t-t_i)^3+\cdots+\frac{1}{50!}\nu^{(50)}
(t-t_i)^{50}.
\end{equation}
The fitted polynomial ($\Phi(t)$, a continuous function) can very
precisely describe the TOA series $\{\Phi(t_i)\}$. One can then take
its first or second derivative to obtain $\nu$ or $\dot\nu$, i.e.
$\nu=\Phi'(t)$ or $\dot\nu=\Phi''(t)$. This procedure is suggested
by the anonymous referee.

\begin{figure*}
\centering
\includegraphics[angle=0,scale=0.6]{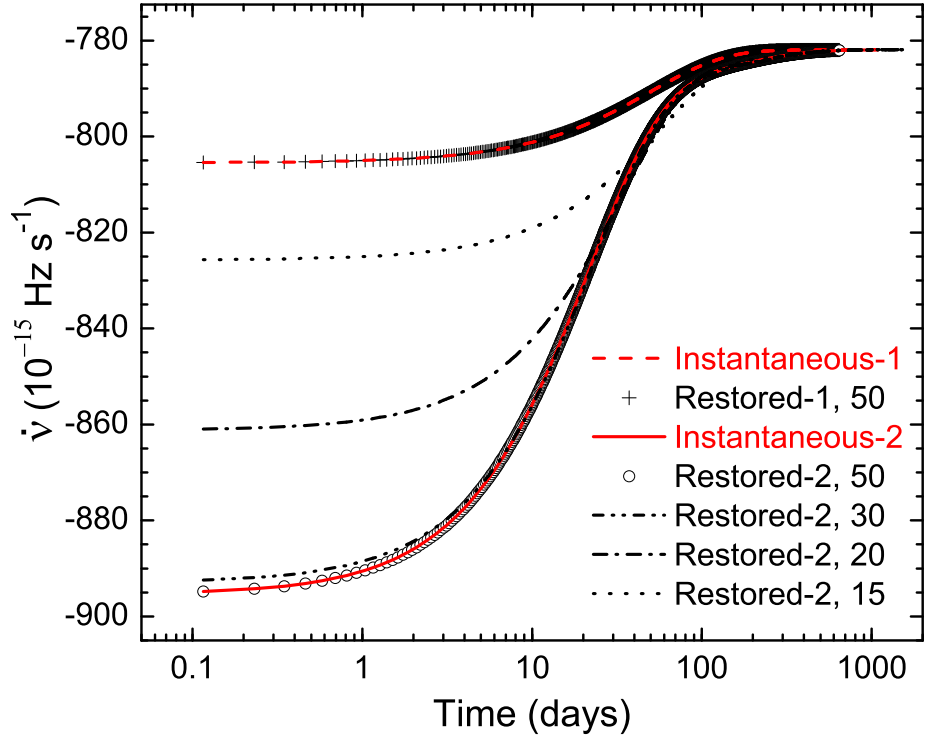}
\caption{Classical glitch simulations with the phenomenological
model. $\dot\nu(t)$ for both one and two component cases (denoted as
1 and 2 in the legend, respectively) are obtained by fitting
procedure IV. Note that restored results are also continuous
functions; here we take their discrete values in order to compare
them more conveniently with the instantaneous values. The numbers
50, 30, 20 and 15 in the legend denote the orders of the fitting
polynomials.} \label{Fig:11}
\end{figure*}

We also simulate the one component and two component cases,
respectively. The results are shown in Figure \ref{Fig:11}. One can
see that the instantaneous values of $\nu(t)$ or $\dot\nu(t)$ can be
restored with very high precision for both the cases. In the figure,
$\Delta T_{\rm int}=10^{6}~{\rm s}$ is taken, and it is checked that
the results are almost independent of both $\Delta T_{\rm int}$ and
$T_{\rm s}$. Then, one can get $\tau_i$ and $\Delta\nu_{{\rm d}i}$
by fitting the restored $\dot\nu(t)$ to Equation
(\ref{relaxation2}). The fitted glitch parameters for one component
case are $\tau=50.00$ days and $\Delta\nu_{\rm
d}=1.01\times10^{-7}~{\rm Hz}$, and for two component case are
$\tau_1=21.37$ days, $\tau_2=146.93$ days, and $\Delta\nu_{\rm
d1}=1.92\times10^{-7}~{\rm Hz}$, $\Delta\nu_{\rm
d2}=1.19\times10^{-7}~{\rm Hz}$. They are all consistent with
instantaneous values (see e.g. Table \ref{Tab:6}) with very high
precisions. In Figure \ref{Fig:11}, we show the fitting results of
two component case with different order polynomials else. It is
found that the order of the polynomial must be very high, e.g.
$\gtrsim 35$, which requires that the TOA data points should not be
too sparse. We also test the fit procedure with different values of
$\tau_i$ and $\Delta\nu_{{\rm d}i}$, and all the glitch parameters
are restored satisfactorily.

In conclusion, procedure III is a reasonable choice to get $\tau$
and $\Delta\nu_{\rm d}$; however, the two components should be fit
simultaneously (in order to avoid some local minimum of $\chi^2$),
and $T_{\rm s}$ should not be too long. Procedure IV seems to be the
best choice for pulsar glitch data analysis, which gives
$\{\nu(t)\}$ and $\{\dot\nu(t)\}$ with very high precision, and then
the glitch parameters $\tau_i$ and $\Delta\nu_{{\rm d}i}$ can be
satisfactorily estimated by fitting the restored $\dot\nu(t)$ to
Equation (\ref{relaxation2}). We thus suggest that theorists should
always use the full timing solution, rather than try to compare
models to individual parameters of fits, as these may be highly
inaccurate. Furthermore working in phase seems to be the most
accurate and reliable method.

\section{Discussions}

\subsection{How to obtain the correct model parameters of pulsars?}

We have shown recently that fitting the observed TOAs of a pulsar to
Equation~(\ref{phase}) will result in biased (i.e., averaged)
spin-down parameters, if its spin-down is non-secular and the
variation time scale is comparable to or shorter than the time span
of the fitting (Zhang \& Xie 2012a, 2012b). In particular we
predicted that the reported braking index should be a function of
time span and approaches to a small and positive value when the time
span is much longer than the oscillation period of its spin-down
process, which can be tested with the existing data (Zhang \& Xie
2012b).

We notice that in some of the literature (e.g. Roy et al. 2012,
Espinoza et al. 2011, Yuan et al. 2010) only Equation~(\ref{phase})
is referred to, when describing the fitting process, even for the
glitch data analysis. However we have shown in Figure~\ref{Fig:0}
that this will produce a significantly distorted glitch profile.
Instead, one can fit to Equations~(\ref{phase2}) and (\ref{phase3})
to obtain the un-distorted (but still averaged) glitch profile;
probably this is usually done in practice, though not explicitly
described in the literature (Yu 2003). We suggest that the exact
fitting procedures should be described when reporting the analysis
results of the observed glitch data.

\begin{figure*}
\centering
\includegraphics[scale=0.55]{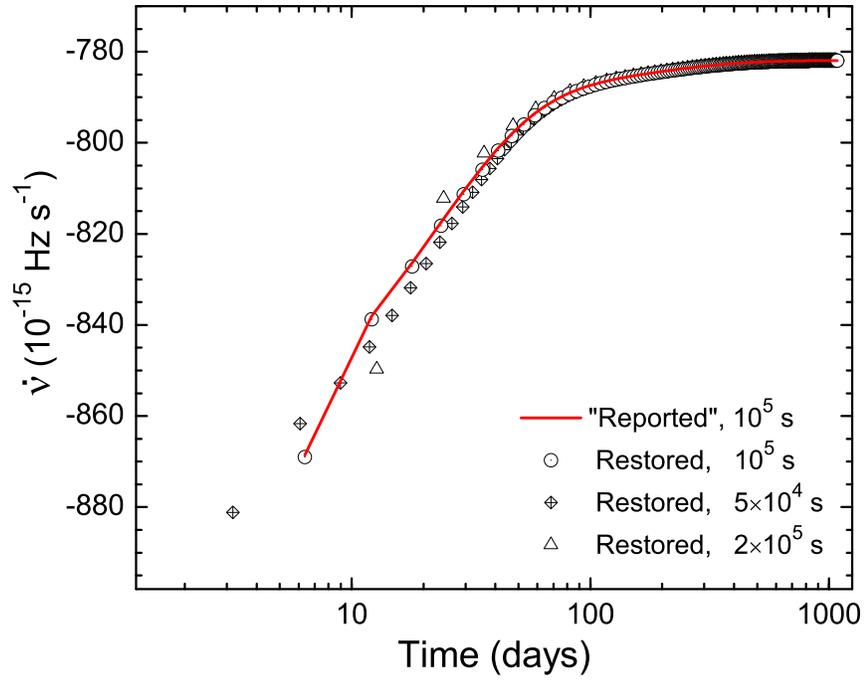}
\caption{The effects of $\Delta T_{\rm int}$ on variation of
$\dot\nu$. The ``reported" $\{\dot\nu\}$ is represented by the solid
line, the restored $\{\dot\nu\}$ with $\Delta T_{\rm int}=10^5$,
$5\times10^4$ and $2\times10^5$ s are represented by circles,
diamonds and triangles, respectively.} \label{Fig:13}
\end{figure*}

However, neither Equation~(\ref{phase}) nor Equations~(\ref{phase2})
and (\ref{phase3}) describe exactly the physical spin-down processes
of all pulsars. The pulsar parameters fitted by
Equations~(\ref{phase4}) are also slightly biased especially if
$T_{\rm s}$ is not properly taken. Ideally, a spin-down model
(empirical, phenomenological or physical) should be used to the fit
the observed TOA data, in order to obtain the model parameters. To
serve this purpose, the observed TOAs of each pulsar should be made
available, and the exact fitting procedure should be described along
with the reported spin-down parameters of a pulsar. As shown in
Figure \ref{Fig:13}, we simulate a glitch recovery with parameters
$\tau_1=21.4$ days, $\tau_2=147$ days, and $\Delta\nu_{\rm
d1}=1.90\times10^{-7}~{\rm Hz}$, $\Delta\nu_{\rm
d2}=1.19\times10^{-7}~{\rm Hz}$, and $\Delta T_{\rm int}=10^5$ s.
Then we have TOAs from the phenomenological model, and the simulated
``reported" $\{\dot\nu\}$ is obtained by fitting TOAs to
Equation~(\ref{phase3}) and is represented by the solid line. By
fitting TOAs to Equation~(\ref{phase4}), we have the ``reported"
glitch parameters $\tau_1=21.8$ days, $\tau_2=152$ days. With the
timescales, we simulate $\{\dot\nu\}$ again with $\Delta T_{\rm
int}=10^5$.  The model parameters $\Delta\nu_{\rm d1}$ and
$\Delta\nu_{\rm d2}$ can be adjusted until simulated fits match the
``reported" ones, and the best fit model parameters $\Delta\nu_{\rm
d1}=1.92\times10^{-7}~{\rm Hz}$, $\Delta\nu_{\rm d2}=1.20\times
10^{-7}~{\rm Hz}$, which are agree well with original parameters. We
show the restored $\{\dot\nu\}$ with circles. With the same
parameters, the restored $\{\dot\nu\}$ for $\Delta T_{\rm
int}=5\times10^4$ and $2\times10^5$ s are represented by diamonds
and triangles, respectively. One can see that $\{\dot\nu\}$ can be
well restored if TOAs are known. If $\Delta T_{\rm int}$ taken in
simulation is not the right one, the $\{\dot\nu\}$ profiles are
apparently different from the ``reported" one, even though the model
parameters are all correct.

When TOAs are available, one can then follow the steps we used above
to combine a model with simulations to obtain model parameters.
Alternatively, $\Phi(t)$, $\nu(t)$ and $\dot\nu(t)$ given by
procedure IV can also be fitted directly by physical models.

\subsection{The effects of discontinuous observations}

In the above analysis, we have assumed that $t_0$ is known; however,
$t_0$ is usually taken as the averaged time of the last reported TOA
just before the glitch and the first reported TOA of the glitch.
This means we have an uncertainty in $t_0$: $\sigma_{t_0}=\Delta
T_{\rm int}/2$. Then from Equation~(\ref{nu_t1}) for a classical
glitch, we find

\begin{equation}\label{sigma}
\frac{\sigma_{\Delta \nu}}{\Delta \nu}=\frac{\sigma_{\Delta
\dot\nu}}{\Delta \dot\nu}=\frac{\sigma_{t_0}}{\tau},
\end{equation}
where $\sigma_{\Delta \nu}$ and $\sigma_{\Delta \dot\nu}$ are the
uncertainties of the restored $\Delta \nu$ and $\Delta \dot\nu$,
respectively. For the classical glitch of B2334+61,
$\sigma_{t_0}\sim 4.2 $ d, $\tau\sim 21.4$ d. Thus from Equation
(\ref{sigma}), we have $\sigma_{\Delta \nu}/\Delta
\nu=\sigma_{\Delta \dot\nu}/\Delta \dot\nu \sim 20\%$.

In principle, we have $\sigma_{\Delta \nu}\approx0$ for a slow
glitch, since $\Delta\nu_{\rm d}$ is determined by the data at the
end of the recovery, i.e. $\nu\sim\Delta\nu_{\rm d}$ for $\Delta
t\gg\tau$ from Equation (\ref{nu_t2}). However, \textbf{from the
derivative of Equation (\ref{nu_t2})}, we have $\dot\nu=\frac{\Delta\nu_{\rm
d}}{\tau} e^{-(t-t_0)/\tau}$ and $\Delta\dot\nu_{\rm
d} (\equiv \Delta\nu_{\rm d}/\tau)$ is closely related to $t_0$,
which resembles the case of a classical glitch. However, for the
slow glitch we can fit for $t_0$ of the glitch by calculating where
the rise and the pre-glitch solutions intersect, which will cause a
much smaller uncertainty. This is a major difference from analyzing
the data of a classical glitch. Unfortunately, this has not been
realized previously and thus $t_0$ was not determined from the
reported $\nu_{\rm O}$ with this method for slow glitch data
analysis. This causes an uncertainty to $\Delta \dot\nu$ in the same
way as in Equation~(\ref{sigma}), i.e., the bias is related to
$\Delta T_{\rm int}$. For instance, in Figure~2 of Zou et al.
(2004), the observed results for a slow glitch event of B1822--09
are $\Delta\nu^a=(40.57\pm26)\ {\rm nHz}$ and $\Delta\dot\nu^a
\simeq 3.1 \times 10^{-15}\ {\rm s}^{-2}$; and for the same event,
the results in Shabanova (2005) are $\Delta\nu^b=40.8~{\rm nHz}$ and
$\Delta\dot\nu^b \simeq 1.4 \times 10^{-15}~\rm s^{-2}$. As expected
above, $\Delta\nu^a=\Delta\nu^b$, but
$\Delta\dot\nu^a\neq\Delta\dot\nu^b$. For the event, $\tau\sim 110$
d, and $\sigma_{t_0}^a\sim 5.5$ d, $\sigma_{t_0}^b\sim 22.8$ d. From
Equation (\ref{sigma}), we obtain
$\sigma_{\Delta\dot\nu}^a=1.6\times10^{-16}~{\rm s}^{-2}$,
$\sigma_{\Delta\dot\nu}^b=6.4\times10^{-16}~{\rm s}^{-2}$, and
$\sigma_{\Delta\dot\nu}=\sqrt{(\sigma_{\Delta\dot\nu}^a)^2+(\sigma_{\Delta\dot\nu}^b)^2}=6.6\times10^{-16}~{\rm
s}^{-2}$. Then we have
$(\Delta\dot\nu^a-\Delta\dot\nu^b)/\sigma_{\Delta\dot\nu}\simeq2.6$,
explaining at least partially the difference between the reported
values of $\sigma_{\Delta\dot\nu}$.

\subsection{Opposite Trends of Recoveries of Slow and Classical
Glitches}

\begin{figure*}
\centering
\includegraphics[scale=0.55]{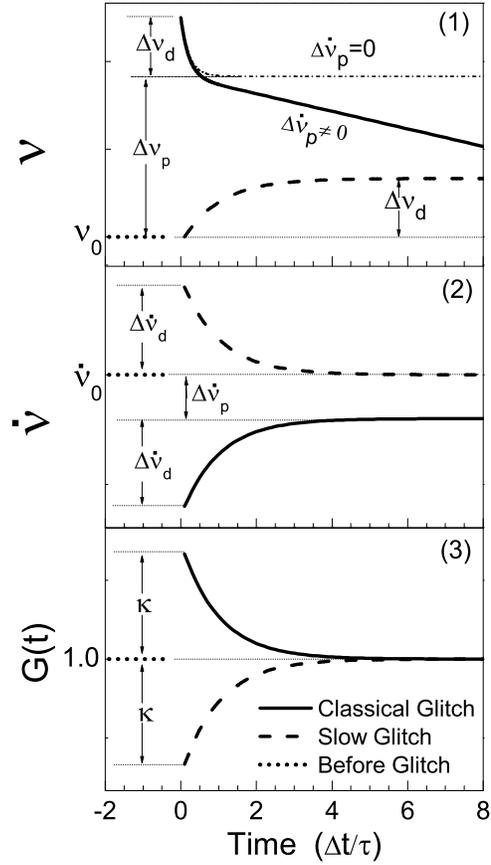}
\caption{Schematic depictions of $\nu$, $\dot\nu$ and $G(t)$ for the
slow and classical glitch recoveries. The pre-glitch tracks are
represented by dotted line. The classical glitch recoveries are
represented by solid lines. The slow glitches are represented by
dashed lines.} \label{Fig:7}
\end{figure*}

Based on observational results, we generalize the variations of
$\nu$ and $\dot\nu$ for slow and classical glitch recoveries, as
shown in Figure~\ref{Fig:7}. The pre-glitch tracks are represented
by dotted line. After the jump, the classical glitch recoveries
(represented by solid line) generally have $\nu$ variation that
tends to restore its initial values, and usually the restoration is
composed by a exponential decay and a permanent linear decrease with
slope $\Delta\dot\nu_{\rm p}$; however, for slow glitches
(represented by dashed line), $\nu$ monotonically increases, as
shown in panel (1). In panel (2), $\dot\nu$ of classical glitch
recoveries that tends to restore its initial values, but cannot
completely recover for $\Delta\dot\nu_{\rm p}\neq 0$; $\dot\nu$ of
slow glitch recoveries almost completely recover to its initial
value, corresponding to the increase of $\nu$.

In sections 3 and 4, we have shown that the classical and slow
glitch recoveries can be well modeled by a simple function, $G(t)=1+
\kappa\exp{(-\Delta t/\tau)}$, with positive or negative $\kappa$,
as shown in panel (3), respectively. However, it is should be
noticed that the model only have two parameters, $\kappa$ and
$\tau$, from which we can obtain $\Delta\nu_{\rm d}$ and
$\Delta\dot\nu_{\rm d}$, but not $\Delta\nu_{\rm p}$ and
$\Delta\dot\nu_{\rm p}$, which are not modelled. Nevertheless, we
conclude that the major difference between slow glitch and classical
glitch recoveries are that they show opposite trends with opposite
signs of $\kappa$, in our phenomenological model.

\section{Summary}

In this work we studied the data analysis procedures of pulsar's
glitch observations and found the conventionally used methods
produce biases to the true glitch parameters with varying degrees.
We presented a phenomenological model for the recovery processes of
classical and slow glitches, which is used to model successfully the
observed slow and classical glitch events from pulsars B1822--09 and
PSR B2334+61, respectively. Based on the model, we tested four
different data analysis procedures. Our main results are summarized
as follows:

\begin{enumerate}

\item The timing analysis method of fitting the observed TOAs with Equation~(\ref{phase2}) or
Equation~(\ref{phase3}) results in significant biases to glitch
parameters of variation magnitude, as shown in Figures~\ref{Fig:1}
and \ref{Fig:2}, and Table \ref{Tab:1}. The biases can be ignored
only when $\Delta T_{\rm int}\le 10^4$~s; otherwise biases still
exist to some extend.

\item With Equation (\ref{phase4}), one can obtain the glitch parameters by fitting the phase
directly, which produce relatively smaller biases. However, for the
case with multiple decay terms, the timescales are usually fixed by
eye for their initial values, which may introduce strong biases.

\item We propose a phenomenological model of glitch recovery (Equation~(\ref{rredipole})), which can reproduce
the commonly observed exponential glitch recovery profiles. The
recovery processes of both slow and classical glitches can be
explained as the $G(t)=1+\kappa\exp{(-\Delta t/\tau)}$ with
$\kappa<0$ (Figure~\ref{Fig:3}) or $\kappa>0$
(Figure~\ref{Fig:4}--\ref{Fig:6}), respectively. Their opposite
trends and main characteristics are illustrated in
Figure~\ref{Fig:7}.

\item Based on the phenomenological model, We simulate four fitting procedures and find that the best one is taking a very high order polynomial
to fit the phase and then taking its derivatives to obtain $\nu(t)$
and $\dot\nu(t)$. Then the glitch parameters can be obtained from
$\nu(t)$ and $\dot\nu(t)$ (e.g. fitting $\dot\nu(t)$ to Equation
(\ref{relaxation2})). We suggest that this procedure should be used
in pulsar timing analysis.

\item The uncertainty in the starting time ($t_0$) of a classical glitch causes uncertainties to the glitch parameters $\Delta\nu_{\rm d}$ and $\Delta\dot\nu_{\rm d}$ (Equation \ref{sigma}), but less so to a slow glitch and $t_0$ of a slow glitch can be determined from data.

\end{enumerate}

However our phenomenological model cannot account for the
non-recoverable jumps in $\nu$ and $\dot\nu$, which are observed for
some classical glitches and may be due to the permanent increase of
a pulsar's dipole magnetic field due to glitches (Lin \& Zhang
2004). In the work, we also assumed uniform TOA distributions to
simulate both the slow and classical glitch recoveries, since the
observed TOAs are not reported in literature. The glitch parameters
can be better restored, if the observed TOAs are available and
fitted directly with a glitch model; this is actually generally
desired for pulsar timing studies. Thus we suggest that TOAs should
be made available to the community when possible or that the full
fitting procedure and fit parameters for different epochs made
available. Also theorists could try to calculate phase as an output,
thus making the comparison more accurate.

\acknowledgments

We thank Jianping Yuan and Meng Yu for valuable discussions. We
would like to thank the anonymous referee for his/her comments and
suggestions that led to a significant improvement of this paper. SNZ
acknowledges partial funding support by 973 Program of China under
grant 2009CB824800, by the National Natural Science Foundation of
China under grant Nos. 11133002 and 10725313, and by the Qianren
start-up grant 292012312D1117210.

\label{lastpage}

\end{document}